\documentclass[a4paper,epsfig]{article}
\usepackage{amssymb}
\usepackage{graphics,graphicx,epsfig}
\usepackage{epstopdf}

\def\nn{\nonumber\\}
\newdimen\Tdim
\def\ispan{{\setbox0=\hbox{i}%
\Tdim\ht0\advance\Tdim\dp0\rule[-\dp0]{0pt}{\Tdim}}}
\def\jspan{{\setbox0=\hbox{j}%
\Tdim\ht0\advance\Tdim\dp0\rule[-\dp0]{0pt}{\Tdim}}}
\def\Tspan#1{{\setbox0=\hbox{#1}%
\Tdim\ht0\advance\Tdim\dp0\advance\Tdim.55ex\rule[-\dp0]{0pt}{\Tdim}\box0}}

\newcommand{\Fymn}[2]{{
\begin{minipage}[h]{#1
\textwidth}
\includegraphics[width=\linewidth,keepaspectratio=true]{#2.eps}
\end{minipage}}}
\makeatletter
\@addtoreset{equation}{section}

\renewcommand{\thefootnote}{\fnsymbol{footnote}}
\makeatother
\def\D{{\cal D}}
\def\p{{\partial}}

\def\gl{{\,
\raisebox{.4em}{\scalebox{1}[.6]{$>$}}\hspace{-.665em}\raisebox{-.04em}{\scalebox{1}[.6]{$<$}}
\, }}
\begin{document}
\thispagestyle{empty}
\begin{flushright}
IFUP-TH/2015\\
July 2015 \\
\end{flushright}
\vspace{3mm}
\begin{center}
{\LARGE  
Small Winding-Number Expansion: \\
Vortex Solutions at Critical Coupling
} \\ 
\vspace{20mm}

{\large
Keisuke~Ohashi$^{a,b,c}$
}
\footnotetext{
e-mail addresses: \tt keisuke084(at)gmail.com, keisuke.ohashi(at)for.unipi.it.
}

\vskip 1.5em
{\footnotesize
{$^a$\it Department of Physics, "E. Fermi", University of Pisa,\\
Largo Pontecorvo, 3,  56127 Pisa, Italy\\
$^b$INFN, Sezione di Pisa,
Largo Pontecorvo, 3,  56127 Pisa, Italy}\\
$^c${\it  Osaka City University Advanced Mathematical Institute (OCAMI),\\
3-3-138 Sugimoto, 
Sumiyoshi-ku, Osaka 558-8585, Japan}
}
 \vspace{12mm}

\abstract{
We study  an axially symmetric solution
of a vortex in the Abelian-Higgs model at critical coupling in detail.  
Here we propose a new idea for a perturbative expansion of a solution, 
 where the winding number of a vortex is naturally extended 
to be a real number and 
the solution is expanded with respect to it  around its origin. We test this idea on three typical
 constants contained in the solution and confirm that  
this expansion works well with the help of the Pad\'e approximation.
For instance, we analytically reproduce the value of the scalar charge of
 the vortex with an error of $O(10^{-6})$. 
This expansion is also powerful even for large winding numbers. 
}

\end{center}

\vfill
\newpage
\setcounter{page}{1}
\setcounter{footnote}{0}
\renewcommand{\thefootnote}{\arabic{footnote}}

\section{Introduction}
A significant feature of many gauge theories is the existence of topological solitons
which may appear when the gauge and/or global symmetries are spontaneously broken.
Monopoles, vortices and domain walls are by now familiar, 
and have found important
applications in vast areas of modern physics, such as 
cosmology, condensed matter physics and particle physics.
From another view point, the topological solitons can be seen as nontrivial solutions
of nonlinear differential equations. 
A direct way to study topological solitons is solving 
such nonlinear equations exactly.
For instance, a beautiful systematic method to construct exact 
solutions for instantons
 has been well-established and is widely known as 
the ADHM construction~\cite{Atiyah:1978ri}. 
This is, however, a special case and
for many other types of solitons numerical calculations are needed to study solutions.

The present work concerns
the so-called Abrikosov-Nielsen-Olesen (ANO) vortex as 
the simplest topological soliton with finite energy 
in the $(1+2)$-dimensional theory.
This vortex appears as a topological defect~\cite{Abrikosov:1956sx} in
Ginzburg-Landau theory~\cite{Ginzburg:1950sr} and 
may be viewed as  a static solution to the equations describing the 1+2
dimensional Abelian Higgs model~\cite{Nielsen:1973cs}.
In this theory all the vortex features 
depend on one dimensionless parameter $\lambda = m_{\rm s}/m_{\rm v}$\footnote{
$\lambda/\sqrt{2}$ is known as the Ginzburg-Landau parameter. }: 
the ratio of the Higgs boson mass $m_{\rm s}$
to the vector boson mass $m_{\rm v}$.
The intervortex force is,
roughly speaking, a superposition of an attractive force caused by the Higgs
boson and a repulsive force caused by the vector boson as seen in a
scalar potential~\cite{Speight:1996px} 
\begin{eqnarray}
 U(R)\simeq \frac{v^2}{2\pi}\left(-q_{\rm s}^2K_0(m_{\rm s} R)+q_m^2 K_0(m_{\rm v} R)\right)
\end{eqnarray}
for a well-separated pair of vortices with a large distance $R$.
Here,  $q_{\rm s}$ and $q_m$ stand for a vortex scalar charge
and  a magnetic dipole moment, respectively. 
Therefore the force with the longest correlation length is dominant 
and  the vortices attract (repel) each other for $\lambda <1$ ($\lambda >
1$)~\cite{Jacobs:1978ch}.
The critical coupling $\lambda =1$ is a rather special case where 
net intervortex forces are exactly canceled thanks to  
the coincidence of the two 
coefficients, $q_{\rm s}=q_{\rm m}\equiv 2\pi C_1$.
From a mathematical viewpoint, 
the Euler-Lagrange equations reduce to the first order differential
equation called 
the Bogomol'nyi-Prasad-Sommerfield (BPS) equations 
for vortices saturating Bogomol'nyi
bound, whose total energy is quantized as $E_k=|k | \pi v^2$  with
the winding number $k\in \mathbb Z$.
In this critical case,  
the constant $C_1$ appears, for instance,  in a potential for
a pair of moving vortices~\cite{Manton:2002wb} 
\begin{eqnarray}
 U_{\lambda=1}(R,\vec u)\simeq \pi v^2\times C_1^2 K_0(m_{\rm v} R)|\vec u|^2+{\cal
  O}(|\vec u|^4)
\end{eqnarray}
with a relative velocity $\vec u$, 
since only the magnetic field accepts a Lorentz boost and the two forces
are not canceled out.
Unlike the remarkable cases of instantons and monopoles, 
no analytic solutions 
for this BPS equation in flat spacetime  have been found 
 even  at this critical coupling. Thus only a few quantities 
are exactly  calculable 
and a detailed study of the vortices, for instance, the calculation of a value
of $C_1$ requires numerical analysis. 

In this paper, to complement the numerical analysis, 
we propose a simple and straightforward, but new
idea for analyzing vortices at critical coupling, where fields are expanded perturbatively with respect to
the winding number $k\in \mathbb Z$ around its origin $k=0$.
To justify this perturbative expansion, 
(let us call it {\it ``small winding-number expansion''}),  
the BPS equations must be extended so that they allow a real winding number 
$k \in \mathbb R$.  Since the BPS equations with
an infinitesimal winding number $|k|\ll 1$ can be exactly solved, 
we can systematically perform perturbation calculations without tuning
any parameters and
this perturbative expansion is supposed to work  well as a practical tool. 
Here, we calculate values of three typical quantities 
with $\lambda=1$ including $C_1$ as the most
simple examples to check this idea. 

The constant $C_1$ has often been calculated in the literature.
De Vega \& Schaposnik \cite{de Vega:1976mi}
gave a semi-analytical study for axially-symmetric solutions 
with an arbitrary winding number $k\in \mathbb Z_{>0}$, 
and constructed power-series expansions around a center of a vortex and 
asymptotic expressions for the opposite side. These two can be 
determined by only one constant $D_k^{k+1}$ for the power-series
expansion 
and $C_k$ ($Z_k$ in their notation) for the asymptotic expression. 
Comparing these parameters in a middle region, 
they obtained  the values: $C_1 = 1.7079...$ and
$D_1^2=0.72791...$\,. 
These values now seems to be widely accepted in literature, for instance,
$C_1 = 1.7079$ appears in Refs.\cite{Cabra:1991eb,Tong:2002rq,GonzalezArroyo:2004xu,Eto:2009wq} and also in a standard textbook of
Vilenkin \& Shellard~\cite{Vilenkin:1994}.
However, we encounter a different value for $C_1$: $C_1 = 10.58/2\pi \simeq 10.57/2\pi \simeq 1.682\sim 1.684$
which was obtained by Speight~\cite{Speight:1996px} about twenty years 
later than de Vega \& Schaposnik~\cite{de Vega:1976mi}.
Furthermore, Tong \cite{Tong:2002rq} gave the supergravity prediction 
$C_1 = 8^{1/4} \simeq 1.68179...$ which seems to agree well with Speight's $C_1$.
These values also seem to be accepted in literature, for instance, 
Ref.\cite{Manton:2002wb} and another standard textbook 
by Manton \& Sutcliffe~\cite{Manton:2004tk}.
There exists a 1.5 \% discrepancy between old and new results.

In Sec.2.6, we shall conclude that the correct value is 
the old one $C_1=1.7079$ by using two different kinds of numerical
calculations with higher accuracy. In Sec.4.2.3, we  reproduce 
this value by using the small winding-number expansion to verify its
power.

This paper is organized as follows.
In Sec.2, we review the BPS vortex in the Abelian-Higgs theory,
and define an extended vortex function which allows
 the winding number of non-integer, as a solution of the BPS equations.
There a non-trivial integral formula including the vortex function is derived 
and three typical constants $C_\nu,D_\nu$ and $S_\nu$ for a
vortex solution are introduced and their analytical and numerical properties are
discussed.
In Sec.3 we perform  a small winding-number expansion of the vortex
function and the three constants  using Feynman-like diagrams. 
Results obtained there are modified in Sec.4, using the Pad\'e approximation to 
overcome problems with finite convergent radii of the expansions.
Summary and discussion are given in Sec.5,  and some useful inequalities
 and  details of the calculations  are summarized in the Appendices.
\section{Review of ANO vortex at  critical coupling}
\subsection{Set up for ANO vortex }

The Abrikosov-Nielsen-Olesen (ANO) vortex is an elementary topological soliton in the
2+1 dimensional Abelian-Higgs model
\begin{eqnarray}
{\cal L} = - \frac{1}{4e^2}F_{\mu\nu}F^{\mu\nu} + \frac12 (\D_\mu
 \phi)^*\D^\mu \phi - V(\phi),
\label{eq:lag_h}
\end{eqnarray}
where $\phi$ is a complex scalar field, metric is $\eta_{\mu\nu} = {\rm diag.}(+1,-1,-1)$ and 
covariant derivative is $\D_\mu = \p_\mu + i A_\mu$.
A scalar potential $V(\phi)$ is  of the wine-bottle type 
\begin{eqnarray}
V(\phi) = \frac{\lambda^2 e^2}{8}\left(|\phi|^2 - v^2\right)^2,
\label{eq:pot}
\end{eqnarray}
which has a vacuum $|\phi|=v$
where the $U(1)$ gauge symmetry is spontaneously broken.
The Higgs mechanism  makes the scalar and the gauge fields massive. 
Their masses are given by, $m_{\rm s} = \lambda \,ev$, 
$m_{\rm v} = ev$ respectively.
The spontaneously broken $U(1)$ symmetry gives rise to a soliton 
which is  topologically stable object supported by $\pi_1(U(1))$, 
of which element is called a winding number.  
To require vanishing of the kinetic term $|{\cal D}_i\phi|^2=0$ at the
spatial infinity 
connects  this winding number with  the first Chern class  
 \begin{eqnarray}
\pi_1(U(1)) = \mathbb{Z}\quad \ni \quad k=-\frac{1}{2\pi}\int d^2x
 F_{12}.
\label{eq:topology}
\end{eqnarray}
This topological defects are called the Abrikosov-Nielsen-Olesen vortex.

In this paper, we take 
the critical coupling constant, $\lambda = 1$, as the simplest
model, 
where the two masses are identical, $m_{\rm v} = m_{\rm s} \equiv m$. Then we can perform
the Bogomol'nyi completion of an energy density $\cal H$ 
for static configurations as
\begin{eqnarray}
{\cal H}\big|_{\lambda=1} 
&=& \frac{1}{2e^2}\left\{F_{12} \pm \frac{e^2}{2}\left(v^2-|\phi|^2\right)\right\}^2
+\frac12 \left|(\D_1 \pm i \D_2)\phi\right|^2\nn
&&\quad  \mp \frac{v^2}2 F_{12}\pm \frac{i}2 \epsilon^{ij}\p_i \left(\phi {\cal D}_j \bar
						\phi\right),
\end{eqnarray}
and a total mass (tension in higher dimension) of vortices, $T$, has a lower bound 
\begin{eqnarray}
 T=\int d^2x {\cal H}\big|_{\lambda=1}
\ge \mp \frac{v^2}2 \int  d^2x F_{12}=\pm \pi v^2 k.
\label{eq:energybound}
\end{eqnarray} 
The inequality is saturated by BPS states which satisfy the BPS equations
\begin{eqnarray}
\mp F_{12} = \frac{e^2}2\left(v^2-|\phi|^2 \right),\quad
(\D_1 \pm i \D_2)\phi = 0.
\label{eq:bps}
\end{eqnarray}
Without loss of generality we will consider the BPS equations with the upper sign.
In order to find general solutions of the BPS equations, it is useful to
solve the second equation in Eq.~(\ref{eq:bps}) at first and, 
it can be solved with the complex coordinate $z=x_1+ix_2$ and
introducing a smooth real function $\psi_{\rm reg}=\psi_{\rm reg}(z,\bar z)$
\begin{eqnarray}
A_{\bar z} = \frac{i}{2}\p_{\bar z}\psi_{\rm reg},\quad
\phi =v\, e^{-\frac{\psi_{\rm reg}}{2}}P(z),  \qquad 
P(z) \equiv \prod_{I=1}^k(z-z_I),
\end{eqnarray}
where an arbitrary holomorphic function $P(z)$ 
can be set to be a monic polynomial without loss of generality.
Here zeros $\{z_I\equiv x^1_I+i x^2_I\in \mathbb C \}$ 
of the Higgs field $\phi$ are topological defects and identified as 
positions of vortices.   One of the important features of BPS vortices is that they feel no interactions
since the attractive and repulsive force are exactly canceled. So we can put
BPS vortices anywhere as many as we like.
Note that the smooth field $\psi_{\rm reg}$ must behave as 
$\psi_{\rm reg}\approx \log|P(z)|^2$ 
at the spatial infinity to obtain a finite
energy, it is convenient and more familiar to rewrite $\psi_{\rm reg}$ 
in terms of a singular field $\psi$,
\begin{eqnarray}
 \psi\equiv \psi_{\rm reg}-\log |P(z)|^2=-\log \frac{|\phi|^2}{v^2},
\end{eqnarray}
so that $\psi$ vanishes at the spatial infinity.
With this singular field, then, the first equation in Eq.~(\ref{eq:bps}) 
can be rewritten to be, so called,  Taubes' equation 
\begin{eqnarray}
-\p_i^2 \psi + m^2 \left(1 - e^{-\psi}\right) =J,
\label{eq:taubes}
\end{eqnarray}
with source terms $J$
\begin{eqnarray}
 J=J(\vec x)=4\pi \sum_{I=1}^k\delta^2(\vec x-\vec x_I).
\end{eqnarray}
Here we used that  the magnetic field can be rewritten as,
\begin{eqnarray}
 -F_{12}=2\p_z\p_{\bar z}\psi_{\rm reg}=\frac12 
\left(\partial_i^2 \psi+J\right) \label{eq:magnetic}
\end{eqnarray}
which coincides with Eq.(\ref{eq:topology}) and $k$ is the total winding number.
Existence and uniqueness of a solution for Taubes' equation with a
 given arbitrary $J$ have been established by~\cite{Taubes:1979tm}. With this
 solution, therefore, we obtain a complete solution for $\phi$ and
 $A_i$.
In terms of a solution of $\psi$ and the source $J$, 
the energy density $\cal H_{\rm BPS}$ for BPS vortices can be rewritten to 
\begin{eqnarray}
 {\cal H}_{\rm BPS}\equiv \frac{v^2}4\left(J+\partial_i^2
				\sigma[\psi]\right),\quad
 \sigma[\psi]\equiv \psi+e^{-\psi}-1\ge 0,\label{eq:BPSenergy}
\end{eqnarray}
which gives the lower bound in Eq.(\ref{eq:energybound}).
There is, however, no known exact solution for this equation,
even in the simplest case with $k=1$.

\subsection{Extension of Taubes' equation
 and particle  description}\label{sec:extension}
In a case that 
$k_I$ vortices coincide at $\vec x=\vec x_I$ for each $I$,   
the source terms are replaced with
\begin{eqnarray}
 J=4\pi\sum_{I} k_I \delta^2(\vec
  x-\vec x_I) ,\quad k=\sum_{I}k_I.
\end{eqnarray}
where $k_I$ indicates the winding number at $\vec x=\vec x_I$.
A request that the winding number $k_I$ is positive integer is 
to give the single-valued Higgs field $\phi$ and Profiles of $\psi$ and 
the magnetic field in Eq.(\ref{eq:magnetic}) and the energy density in Eq.(\ref{eq:BPSenergy}) 
can be calculated without constructing  $\phi$.  
If we omit constructing $\phi$, therefore, we can formally extend Taubes'
equation with the generalized source terms
\begin{eqnarray}
 J=4\pi \sum_I \nu_I \delta^2(\vec x-\vec x_I),  
\quad \nu_I \in \{\nu\, |\nu>-1,\nu \in \mathbb R\} \label{eq:generalsource}.
\end{eqnarray} 
Here the winding number $k_I$ is renamed $\nu_I$ to stress that $\nu_I$
can be non-integer and the lower bound of the winding numbers will be
discussed in Sec.2.4. 
A `total mass' of this extended object is formally calculated as 
\begin{eqnarray}
 T_{\rm BPS}
=\int d^2x {\cal H_{\rm BPS}}=\pi v^2\times \nu,\quad \nu\equiv
\sum_{I}\nu_I, 
\end{eqnarray}
which takes a negative value for $\nu<0$.
Integrating the both sides of Taubes' equation Eq.(\ref{eq:taubes}) 
we find the following identity corresponding to Eq.(\ref{eq:topology})
\begin{eqnarray}
 \nu=\frac1{4\pi}\int d^2x \left(\partial_i^2 \psi+J\right)=
\frac{m^2}{4\pi}\int d^2x (1-e^{-\psi}),\label{eq:nu}
\end{eqnarray}
which is no longer an element of $\pi_1(U(1))$.
In the rest of this paper, we will study this extended Taubes' equation
with a generalized source term Eq.(\ref{eq:generalsource}) 
and its solution numerically and analytically.  
This extension allows us to consider 
a {\it Taylor expansion of the solution with
respect to the winding numbers} as discussed in Sec.\ref{sec:TCE}, although we are not specially interested in 
the solution with the winding numbers of non-integer. 

Uniqueness of the solution for this extended Taubes' equation 
can be easily shown as Appendix \ref{sec:Uniquness}.  For instance, 
we know  the trivial solution
\begin{eqnarray}
 \psi =0   \quad {\rm for~}\quad J=0.\label{eq:vanishing}
\end{eqnarray} 
To show existence of the solution for the extended Taubes' equation is
difficult and  out of scope of this paper, 
and we just assume the existence of the solution here. 
Therefore, the solution of $\psi$ is a function with respect to 
a coordinate $\vec x$, 
positions of vortices $\{\vec x_I\}$ and their winding number
$\{\nu_I\}$, $\psi=\psi(\vec x,\{\vec x_I,\nu_I\})$.
Furthermore we assume that the solution is differentiable with respect 
to $\{\nu_I\}$. Under this assumption,
we can derive, for each $I$, 
\begin{eqnarray}
 (-\partial_i^2+m^2 e^{-\psi})
\frac{\partial \psi}{\partial \nu_I }
 =4\pi \delta^2(\vec x-\vec x_I)
\end{eqnarray} 
from Taubes equation Eq.(\ref{eq:taubes}) with the source 
Eq.(\ref{eq:generalsource}). 
According to Appendix \ref{sec:Uniquness} the above equation show that 
the solution $\psi$ is strictly increasing with respect to each $\nu_I$, $\p
\psi/\p \nu_I>0$. In the limit of the vanishing source $J=0$,
furthermore we find 
\begin{eqnarray} 
\lim_{J\to 0}\frac{\partial \psi}{\partial \nu_I}=
\frac{4\pi}{-\partial^2_i+m^2}\delta^2(\vec x-\vec x_I)
= 2 K_0(m |\vec x-\vec x_I|),\qquad \label{eq:dpsidnu0}
\end{eqnarray} 
where 
the modified Bessel function of the second kind $K_0(x)$ emerges as a
two-dimensional Green's function. That is, in this limit 
a vortex solution is exactly solved and treated as a linear combination 
of free massive particles and  for small $|\nu_I|\ll 1$ at least, 
$\psi$ is approximated well {\it everywhere} as 
\begin{eqnarray}
 \psi \approx  2\sum_{I}\nu_I K_0(m|\vec x-\vec x_I|).\label{eq:E1general}
\end{eqnarray}
This is the starting point of the small winding-number expansion  which
will be discussed in Sec.\ref{sec:TCE}.

In this particle description, it will be convenient to 
rewrite  Taubes' equation as
\begin{eqnarray}
- \partial_i^2 \psi +m^2 \psi=J +m^2 
\sigma[\psi],\qquad \sigma[\psi]=\psi^2 \sum_{n=0}^\infty 
\frac{(-1)^{n}}{(n+2)!}\psi^n,
\end{eqnarray}
with $\sigma[\psi]$ as dimensionless self-interaction terms.
Then, by applying the Green's function method to Taubes' equation,  
we obtain\footnote{Here we used the fact that
$\psi$ vanishes at the spatial infinity.}
 an integral equation for $\psi$ 
 with  Green's function $G(\vec x)=K_0(m|\vec x|)$,
\begin{eqnarray}
 \psi(\vec x)
&=&2\sum_{I}\nu_I G(\vec x-\vec x_I)+m^2 \int \frac{d^2 y}{2\pi} G(\vec x-\vec y)
\sigma[\psi(\vec y)]. \label{eq:psiIntegral}
\end{eqnarray}
Since $\sigma[\psi]\ge 0$ and $K_0(x)>0$ are always hold, we find that 
the solution of Taubes' equation must 
 satisfy a fundamental inequality 
\begin{eqnarray}
 \psi(\vec x) >2\sum_{I}\nu_I K_0(m|\vec x-\vec x_I|).   
\end{eqnarray}
\subsection{Scaling argument and a physical size of a vortex}
Let us consider the following Lagrangian in a two-dimensional Euclidean
spacetime 
\begin{eqnarray}
 {\cal L}_{\rm BPS}=-\frac12 (\partial_i \psi)^2-
m^2(\psi+e^{-\psi}-1)+J \psi, \quad 
\label{eq:vortexLagrangian}
\end{eqnarray}
which induces  Taubes' equation as an equation of motion of $\psi$, and 
an action\footnote{
Substituting the solution, $K$ becomes a function with respect to
 complex coordinates
$z_I=x^1_I+i x^2_I$ describing positions of vortices.  
With a limit of $m_0\to 0$, 
this quantity gives a K\"ahler potential describing the vortex moduli
space \cite{Chen:2004xu} as
\begin{eqnarray}
 \pi v^2 \sum_{I}\nu_I |z_I|^2+v^2 \lim_{m_0\to 0}K.
\end{eqnarray} 
At the limit $m_0\to 0$, $1/m_0$ gives a IR cut-off and $K_{\rm ghost}$
can be eliminated by K\"ahler transformations. Actually 
one can confirm that the above K\"ahler potential gives  
Samols' metric\cite{Samols:1991ne}.} is 
\begin{eqnarray}
 K=-\int d^2x {\cal L}_{\rm BPS}\Big|_{\rm solution}+K_{\rm ghost},
\label{eq:KEnergy}
\end{eqnarray}
where $K_{\rm ghost}$ is
 introduced to cancel UV divergences of the kinetic term and the
 source term  
and we set $K_{\rm ghost}$ as, for instance, 
\begin{eqnarray}
K_{\rm ghost}=-\int d^2x \left(\frac12 (\partial_i \varphi)^2+\frac12 m_0^2
	       \varphi^2 -J \varphi \right),\quad \nn
 \varphi(\vec x)=2\sum_I \nu_I K_{0}(m_0 |\vec x-\vec x_I|).
\end{eqnarray}
After this regularization we can apply the scaling argument to this action.
For simplicity, let us consider an axially symmetric case with the
source $J=4\pi\nu \delta^2(\vec x)$.
Since $K$ is a dimensionless quantity, 
the dimensional argument tells us 
\begin{eqnarray}
0=m^2 \frac{\partial K}{\partial m^2}+
m_0^2 \frac{\partial K}{\partial m_0^2} 
\label{eq:KDerrick}
\end{eqnarray}
By using equations of motion for $\psi$ and $\varphi$,  derivatives of
$K$  with respect to masses can be calculated by
\begin{eqnarray}
m^2 \frac{\partial K}{\partial m^2}
&=&\int d^2x m^2\left(\psi+e^{-\psi}-1\right)
= m^2 \int d^2x \psi-4\pi \,\nu ,\nn
m_0^2 \frac{\partial K}{\partial m_0^2}&=&
m_0^2 \frac{\partial K_{\rm ghost}}{\partial m_0^2}=
-\int d^2x \frac{m_0^2}2 \varphi^2=-2\pi \nu^2,\label{eq:dKdm}
\end{eqnarray}
where we used Eq.(\ref{eq:nu}).
Therefore, we find the following  formula~\cite{Manton:1998xv}   
\begin{eqnarray}
 \int d^2x\, \psi =\frac{2\pi }{m^2}\times \nu(\nu+2).\label{eq:Derrickformula}
\end{eqnarray}
As we seen the above this exact formula does not come from
topological argument, but from the scaling argument.  
To check numerical calculations we use this formula  in this paper. 
Thanks to this non-trivial identity combining Eq.(\ref{eq:nu}), the following  integral
 is calculated  as  
\begin{eqnarray}
\int d^2x |\vec x|^2 \frac{\cal H_{\rm BPS}}{\pi v^2}=\frac1{8\pi} 
\int d^2x |\vec x|^2\partial_i^2\sigma[\psi]
=\int \frac{d^2x}{2\pi} (\psi+e^{-\psi}-1)=\frac{\nu^2}{m^2},\quad
\label{eq:nusquared} 
\end{eqnarray}
and a size of the vortex with the positive winding number $\nu>0$ can be
naturally defined with the energy density $\cal H_{\rm BPS}$ given in
Eq.(\ref{eq:BPSenergy}) and calculated as,    
\begin{eqnarray}
 R_{\rm BPS}\equiv 
\sqrt{2\times \frac{\int d^2x |\vec x|^2{\cal H_{\rm BPS}}}{\int d^2x
{\cal H}_{\rm BPS}}}
=\frac{2\sqrt{\nu}}{m}\label{eq:Rflux}\quad {\rm for~}\nu>0,
\end{eqnarray}
which turns out to be a key point in Sec.\ref{sec:PadeandLarge}. 
It is natural for the scaling argument to determine 
a typical size of a soliton. 
\subsection{Axially symmetric solution}\label{sec:axial}
\begin{figure}[h]
 \begin{center}
  \includegraphics[width=5.5cm]{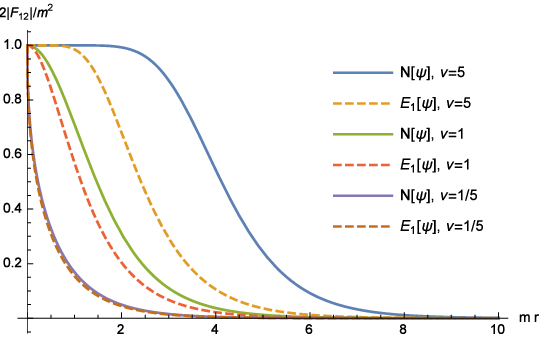}\qquad
  \includegraphics[width=5.5cm]{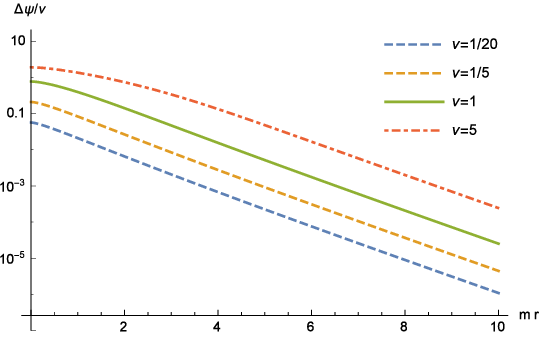}
 \end{center}
\caption{\footnotesize 
Magnetic flux in the left panel and differences 
$\Delta \psi=N[\psi]-E_1[\psi]$ in the right panel 
}\label{fig:psinu}
\end{figure}

Let us consider a single vortex sitting the origin  
with the winding number $\nu$, that is, 
we consider a solution with the source term $J=4\pi \nu \delta^2(\vec x)$.
Its configuration 
is axially symmetric and described by a function $\psi=\psi(m r,\nu)$ 
with respect to a radial coordinate $r=|\vec x|$
and the winding number $\nu$. The partial differential equation (\ref{eq:taubes}), therefore, reduces
to  an ordinary differential equation 
\begin{eqnarray}
 \frac1r \frac{d}{dr}\left(r \frac{d\psi}{dr}\right)=m^2(1-e^{-\psi})
\label{eq:taubes1dim}
\end{eqnarray}
with the following two boundary conditions  
\begin{eqnarray}
 \lim_{r\to 0}r \frac{d \psi}{dr} =-2\nu, \quad  \lim_{r\to \infty
  }\psi=0.
\label{eq:boundcon}
\end{eqnarray}
Even for the non-integer number $\nu$, 
a set of the differential equation and the boundary conditions defines 
 an unique solution under assumption of its existence.  
Especially 
for small $|\nu|\ll 1$, 
$\psi$ is approximated  {\it in the full range} of $r\in \mathbb R_{>0}$ 
as 
\begin{eqnarray}
 \psi \approx E_1[\psi]\equiv 2\nu K_0(mr),\qquad 
\lim_{\nu\to 0}\frac{\psi}{\nu}=\lim_{\nu\to 0}\frac{\p \psi}{\p \nu}=
\psi_1\equiv 2K_0(mr).\label{eq:E1}
\end{eqnarray} 
See Fig.\ref{fig:psinu} for some examples of profile functions of $N[\psi]$ 
which denotes $\psi$ calculated  numerically. 
Here we assume that the solution $\psi$ is smooth with respect to $\nu$ at 
$\nu=0$. This assumption requires the solution to be extended for the 
negative winding number $\nu$.
Since  $\p \psi/\p \nu>0$ as discussed in Sec.\ref{sec:extension}, 
a lower bound of $\nu$ is shown by taking a derivative of the both sides
of Eq.(\ref{eq:Derrickformula}) as
\begin{eqnarray}
0<\int d^2x \frac{\p \psi}{\p \nu}=\frac{4\pi}{m^2}(\nu+1),\label{eq:absence}
\end{eqnarray}
that is, {\it there exist no solution of Taubes' equation with $\nu\le -1$}. 
We just assume the existence of the solution with $\nu>-1$ in this paper.

Note that we can show the following inequalities although we have no
exact solution. Applying the discussion in Appendix.\ref{sec:Uniquness}
to Taubes' equation with the source $J$ in Eq.(\ref{eq:generalsource}), 
we find the solution $\psi$ must be positive for $\nu>0$ and 
be negative for $\nu<0$, and Eq.(\ref{eq:taubes1dim})
tells us 
that $r \frac{d\psi}{dr}$ is strictly increasing (decreasing) with respect
to $r$ for $\nu>0$ ($\nu<0$), 
and therefore the boundary conditions Eq.(\ref{eq:boundcon})
give lower and upper bounds as,
\begin{eqnarray}
 \psi >0,\quad  -2\nu <r \frac{d \psi }{d r}<0,\quad 
&{\rm for~} \nu>0,\nn
\psi<0,\quad -2\nu >r \frac{d \psi }{d r}>0,\quad 
&{\rm for~} -1<\nu<0. \label{eq:psiinequality}
\end{eqnarray}
According to  Appendix \ref{sec:Uniquness}, 
the following inequality 
\begin{eqnarray}
 (-\partial_i^2+m^2 e^{-\psi})
\frac{\partial^2 \psi}{\partial \nu^2 }
=m^2 e^{-\psi} \left(\frac{\partial \psi}{\partial
		   \nu}\right)^2>0,
\end{eqnarray}
implies that $\psi$ is a downward-convex function,
\begin{eqnarray}
\frac{\partial^2\psi}{\partial \nu^2}=
\frac1{\nu}\frac{\p}{\p \nu}\left(\nu \frac{\p \psi}{\p \nu}-
			     \psi\right)>0.\quad \label{eq:ddpsi}
\end{eqnarray}
Combining  Eq.(\ref{eq:E1}) with this fact, we find that $\psi/\nu$ is strictly 
increasing with respect to $\nu$ and furthermore we obtain
\begin{eqnarray}
\frac{\partial \psi}{\partial \nu}>\frac{\psi}{\nu}>2 K_0(m r)>0  \quad
 &{\rm for~} \nu>0,\nn
0<\frac{\partial \psi}{\partial \nu}<\frac{\psi}{\nu}<2 K_0(m r)  \quad
 &{\rm for~} -1<\nu<0.
\label{eq:psiinequality2}
\end{eqnarray}

With this  axially-symmetric  solution 
$\psi(\vec x)=\psi(r)$ with $r=|\vec x|$,
the integral equation Eq.(\ref{eq:psiIntegral}) 
reduces to 
\begin{eqnarray}
 \psi(r)&=&2\nu K_0(mr)
+m^2 \int_0^\infty d s \,s G_{\rm F}(r,s) \sigma[\psi(s)],
\label{eq:psiIntegral2}
\end{eqnarray}
where the reduced Green's function $G_{\rm F}(r,s)$ takes the following form 
\begin{eqnarray}
G_{\rm F}(r,s)&=&\int \frac{d \theta }{2\pi} K_0\left(m\sqrt{r^2+s^2-2 r s \cos\theta
				}\right)\nn
&=& \Theta(r-s) K_0(m r) I_0(m s)+\Theta(s-r)K_0(m s)I_0(m r)
\end{eqnarray}
with the step function $\Theta(x)$ and the modified Bessel function of the
first kind $I_0(x)$. 

\subsection{Observable parameters, $C_\nu,D_\nu,S_\nu$}

\subsubsection{ $D_\nu$  and Internal size
   $R_{\rm in}$}
\begin{figure}[h]
 \begin{center}
   \includegraphics[width=5.5cm]{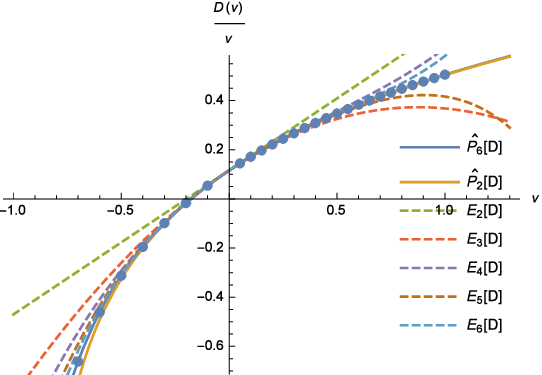}\qquad
  \includegraphics[width=5.5cm]{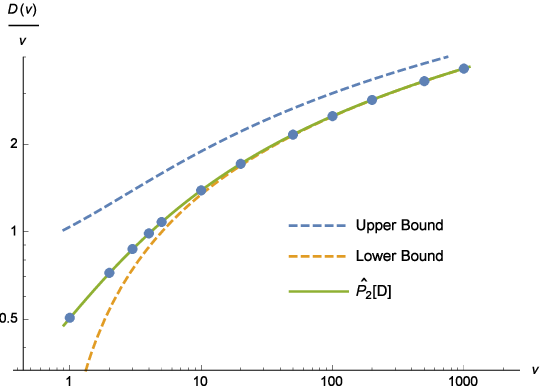}
 \end{center}
\caption{\footnotesize 
Profile of $D_\nu$ for the full range of $\nu$.
Numerical Data $N_{\rm sht}[D_\nu]$ are plotted  by  dots. 
  Dashed lines in the left panels describe $E_n[D_\nu]$ 
given in Sec.\ref{sec:TCE}. Dashed lines in the right panel 
give the bounds given in Eq.(\ref{eq:Rbounds}). 
$\widehat P_2[D_\nu],\widehat P_6[D_\nu]$ 
plotted by a solid line  are defined 
in Sec.\ref{sec:PadeandLarge}.}\label{fig:Dnu}
\end{figure}

To define 
the solution  $\psi$ of Taubes' equation even with 
the positive non-integer winding number $\nu$,
we have to consider a behavior of the solution around 
the core of the vortex seriously.  
Note that in the massless limit $m\to 0$, Taubes' equation 
has a general solution\footnote{Here we omit the boundary condition for
the spatial infinity.} with a 
positive real arbitrary constant $R_{\rm in}$ as,
\begin{eqnarray}
\lim_{m\to 0} \psi = -\log Y,\quad    Y\equiv \left(\frac{r}{R_{\rm in}}\right)^{2\nu} \label{eq:r0}
\end{eqnarray} 
and with the finite mass $m>0$, therefore, $\psi$ can be expanded by $m$
and 
 we find an expansion of $\psi$ around 
the origin $r=0$ in an unfamiliar form,
\begin{eqnarray}
 \psi &=&-\log Y+\sum_{n=1}^\infty F_n(Y) (m r)^{2n}\nn
&\approx & 
-2\nu \log (m r)+2 D_\nu+\left\{ 
\begin{array}{cc}
\frac14(mr)^2 &  {\rm for~} \nu>0\vspace{0.5em}\\ 
-\frac14\frac{e^{-2D_\nu}}{(1+\nu)^2}(m r)^{2(1+\nu)} & {\rm for~} -1<\nu<0
\end{array}\right., \qquad \quad\label{eq:rexp}
\end{eqnarray}
where we treated $mr$ and $Y$ as if they were independent of each
other,
and a function $F_n(Y)$ is independent of $m$ and 
turns out to be  a polynomial of order $n$ with respect to
$Y$ determined sequentially by solving Taubes' equation as,
\begin{eqnarray}
 F_1(Y)=\frac14\left(1-\frac{Y}{(1+\nu)^2}\right),\quad F_2(Y)=
\frac{Y}{64}\left(\frac{4}{(2+\nu)^2}-\frac{Y}{(1+\nu)^2}\right),\cdots 
\end{eqnarray}   
which must vanish in the limit $\nu \to 0$ for a finite radius $r$ 
due to Eq.(\ref{eq:vanishing}).
The  dimensionless constant $D_\nu$ appeared in the expansion 
is related to $R_{\rm in}$ 
as\footnote{
A relation between $D_\nu$ for $\nu=k\in \mathbb Z_{>0}$ and 
$D_k^{k+1}$ defined by de Vega \& Schaposnik~\cite{de Vega:1976mi} is 
\begin{eqnarray}
 D_k^{k+1}=\frac{4^k}{k+1}\exp(-2 D_k),
\end{eqnarray}
For instance, we numerically obtain 
$D_1^2=2\exp(-2\times 0.505360825\dots)=0.72791247\dots$ 
which coincides with their value $D_1^2=0.72791$.
}
\begin{eqnarray}
 D_\nu=\nu \log (mR_{\rm in}).
\end{eqnarray} 
Therefore the expansion of $\psi$ can be defined by a pair of parameters 
$\{\nu, R_{\rm in}\}$. The uniqueness of the solution with a given $\nu$
means, however, that
 to satisfy the boundary condition at the spatial infinity, the constant 
$R_{\rm in}$ must take a certain value corresponding to each value of
$\nu$, that is, a function $R_{\rm in}=R_{\rm in }(\nu)$, 
otherwise a function defined by the expansion always glows up at a large $r$. 
In Appendix \ref{sec:Inequalities} this feature is analytically  discussed
and at the present we find a pair of lower and upper bounds of $R_{\rm in}$ as 
\begin{eqnarray}
\frac{2\sqrt{\nu+1}}{m}>
R_{\rm in}> \frac{2}{m} \sqrt{\frac{\nu}e}\qquad {\rm for~} \nu>0.
\label{eq:Rbounds}
\end{eqnarray}
According to Eq.(\ref{eq:psiinequality2}) 
 $R_{\rm in}$ and $D_{\nu}/\nu $ turn out to be strictly increasing functions with respect to
 $\nu$ and take values at $\nu=0$ 
\begin{eqnarray}
\lim_{\nu\to 0} \frac{D_\nu}{\nu}&=&\lim_{r\to 0}(K_0(r)+\log r)=\log 2-\gamma
\approx 0.115932,\nn
\lim_{\nu\to 0}R_{\rm in}&=&\frac{2e^{-\gamma}}m\approx\frac{1.12292}m,\qquad
\end{eqnarray}
with Euler's gamma $\gamma$.
In Fig.\ref{fig:Dnu},  we plot a profile of $D_\nu/\nu$.
Note that there is an another way to calculate  $D_\nu$ 
using  the integral form Eq.(\ref{eq:psiIntegral2}) as,
\begin{eqnarray}
D_\nu=\lim_{r\to 0}\left(\frac{\psi}2+\nu \log (mr)\right)=\nu(\log 2-\gamma)+\frac{m^2}2 \int_0^\infty ds s
K_0(ms)\sigma[\psi(s)].
\label{eq:Dintegral}
\end{eqnarray}
These different two definitions of $D_\nu$ will be used to double-check
numerical calculations of $D_\nu$.

Since the axially symmetric vortex solution we consider 
has the only one mass parameter $m$, 
we expect that the dimensionfull parameter $R_{\rm in}$ controlling 
a profile of the solution should be the same order of the vortex size
$R_{\rm BPS}$ given in Eq.(\ref{eq:Rflux}).
Thanks to Eq.(\ref{eq:Rbounds}), roughly speaking, 
we find actually $R_{\rm BPS} \approx R_{\rm in}$
for large $\nu$. We call $R_{\rm in}$ an internal size.    
On the other hand 
 $D_\nu$ is directly related 
to a value of the action $K$ with $J=4\pi \nu \delta^2(\vec x)$ in the previous subsection.
In the same way of Eq.(\ref{eq:dKdm}), we can 
calculate a derivative of $K$ with
respective to $\nu$,
\begin{eqnarray}
\frac{\partial K}{\partial \nu}=-4\pi \times \lim_{r \to
 0}\left(\psi-\varphi\right)=8\pi \nu \log\left(\frac{m}{m_0}\right)
-8\pi\left(D_\nu-\nu(\log 2-\gamma) \right)
\end{eqnarray}
and by setting 
the mass of the ghost $m_0$ to be $m_0=2e^{-\gamma }m$, we obtain  
the following simple relations, 
\begin{eqnarray}
  D_\nu=-\frac1{8\pi}\frac{d K}{d \nu},\quad K=-8\pi \int_0^\nu dy D_y.
\end{eqnarray}
\subsubsection{ Scalar charge  $C_\nu$}
\begin{figure}[h]
 \begin{center}
  \includegraphics[width=5cm]{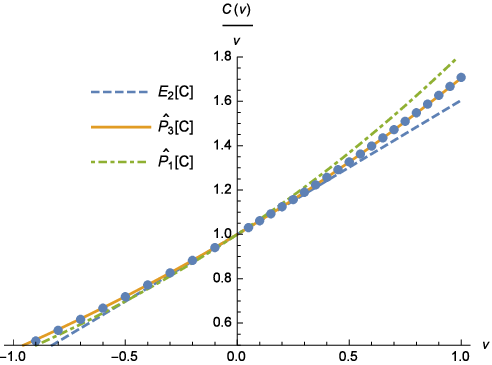}\qquad
  \includegraphics[width=6cm]{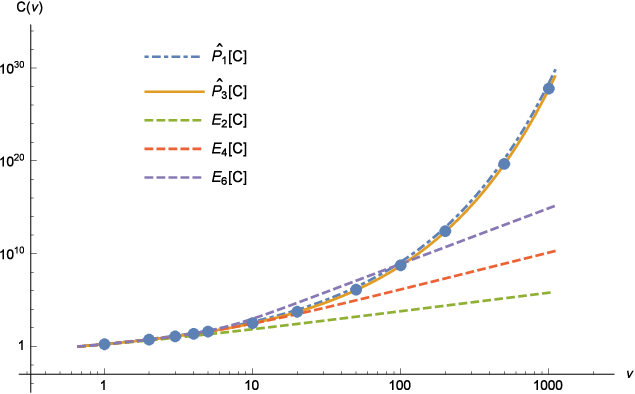}
 \end{center}
\caption{\footnotesize 
Profile of $C_\nu$ for small $\nu$ in the left panel and for large $\nu$
 in the right panel.  
Numerical Data $N_{\rm sht}[C_\nu]$ are plotted  by  dots. 
  Dashed lines in the both panels describe
approximants of the order $n$, $E_n[C_\nu]$, in terms of 
the winding-number expansion discussed in Sec.\ref{sec:TCE}. 
$\widehat P_3[C_\nu]$ plotted by a solid line and  
$\widehat P_1[C_\nu]$ plotted by a dot-dash-line are defined 
in Sec.\ref{sec:PadeandLarge}.
}\label{fig:Cnu}
\end{figure}

Let us take $m$ large conversely, that is, consider a infrared region
$r\gg R_{\rm in}\approx 2\sqrt{\nu}/m$. 
There, an asymptotic behavior of $\psi$ can be treated as a
 fluctuation of a free massive scalar field around the vacuum. 
Due to the axial symmetry, such a fluctuation is written  with a certain
 constant $C_\nu\in \mathbb R_{>0}$ as   
\begin{eqnarray}
 \psi \approx  2 C_\nu K_0(mr). \label{eq:asymptotic}
\end{eqnarray}
There is the similarity between this asymptotic form and
the form of Eq.(\ref{eq:E1}) and the uniqueness of the solution of 
Taubes' equation 
indicates that the two constants $C_\nu$ and $\nu$
 are in one-to-one correspondence.     
Actually, to satisfy the boundary condition at the origin $r=0$, the constant 
$C_\nu$ must be a function with respect to $\nu$
and according to
Eq.(\ref{eq:vanishing}), Eq(\ref{eq:dpsidnu0}) and Eq(\ref{eq:ddpsi}) we find
\begin{eqnarray}
 \lim_{\nu\to 0}C_\nu=0,\quad \lim_{\nu\to 0}\frac{d
  C_\nu}{d \nu}=1,\quad \frac{d^2 C_\nu }{d \nu^2}>0.
\end{eqnarray}
These property tell us that 
$C_\nu/\nu$ is  strictly increasing with respect to $\nu$ and 
a lower bound of $C_\nu$ is given as  $C_\nu>\nu$. 
A profile of this function is shown in Fig.\ref{fig:Cnu}.
According to the integral equation Eq.(\ref{eq:psiIntegral2}), 
$C_\nu$ can be calculated by
\begin{eqnarray}
  C_\nu=\lim_{r\to \infty}\frac{\psi}{2K_0(mr)}=\nu+\frac{m^2}2 \int_0^\infty ds s I_0(m s)\sigma[\psi(s)].
\label{eq:Cintegral}
\end{eqnarray}
Bringing this identity back, 
we can remove the explicit $\nu$-dependence from the integral equation 
Eq.(\ref{eq:psiIntegral2}) as
\begin{eqnarray}
 \psi(\vec x)=2C_\nu K_0(mr)-\int_0^\infty ds\, s G_{\rm ad}(r,s)
  \sigma[\psi(s)],\label{eq:psiIntAdvanced}
\end{eqnarray}
with  an `advanced' Green's function\footnote{
Positivity of this quantity is easily shown since 
$K_0(r)$ ($I_0(r)$) is strictly decreasing (increasing) with respect to $r$. 
  } 
\begin{eqnarray}
G_{\rm ad}(r,s)
  =\Theta(s-r)\left\{K_0(mr)I_0(ms)-I_0(mr)K_0(ms)\right\} \ge 0.
\end{eqnarray} 
Using this integral equation Eq.(\ref{eq:psiIntAdvanced}), the asymptotic
behavior in Eq.(\ref{eq:asymptotic}) is modified as
\begin{eqnarray}
 \psi=2C_\nu K_0(mr)-2C_\nu^2\int_0^\infty ds\,s G_{\rm ad}(r,s)K_0(m s)^2
+{\cal  O}(e^{-3mr}).\label{eq:asymptoticexp}
\end{eqnarray}
Thanks to these two different forms of the integral equations for $\psi$ 
Eq.(\ref{eq:psiIntegral2}) and Eq.(\ref{eq:psiIntAdvanced}),  
we find lower  and upper bounds as
\begin{eqnarray}
 2\nu K_0(mr)<\psi<2C_\nu K_0(mr).
\end{eqnarray}

A one of purposes of this paper is to confirm the true value of $C_1$.

\subsubsection{Total scalar potential $S_\nu$}
Finally let us consider  the following definite integral\footnote{
This quantity also appeared as 
a fundamental constant, $c=2S_1\approx 0.830707$,in Eq.(5.2) 
of a paper \cite{Eto:2012qda}.}
\begin{eqnarray}
 S_\nu=\frac{ m^2}{2} \int \frac{d^2x}{2\pi} (1-e^{-\psi})^2,
\end{eqnarray}
which is dimensionless and proportional to a total potential energy of
the Abelian-Higgs model at critical coupling,
\begin{eqnarray}
  S_\nu&=&\frac{\lambda}{E_1}\frac{\p E_\nu}{\p \lambda} 
\Big|_{\lambda=1}
=\frac2{E_1}\int d^2x
  V(\phi)\Big|_{\lambda =1,\rm sol},
\end{eqnarray}

\begin{figure}[h]
 \begin{center}
  \includegraphics[width=5.5cm]{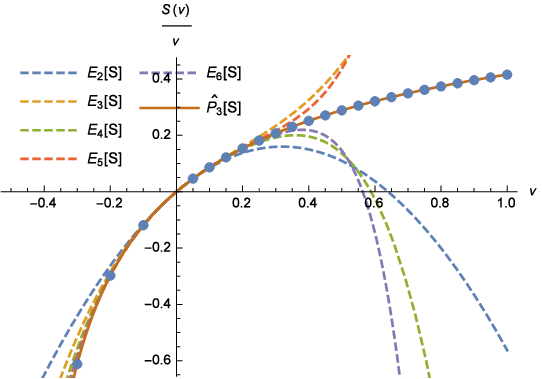}\qquad
  \includegraphics[width=5.5cm]{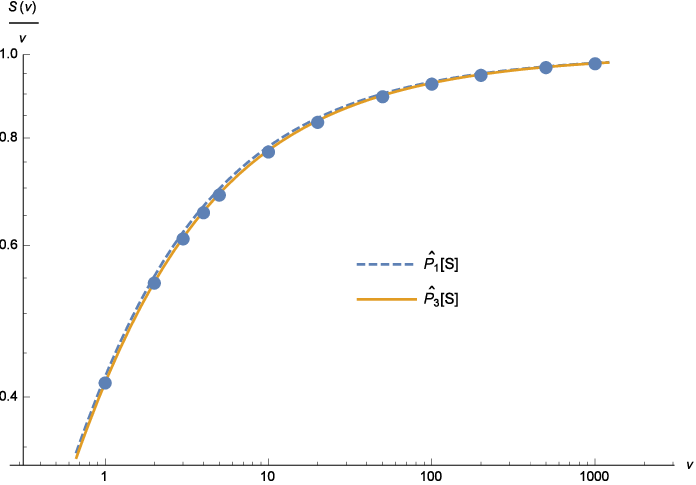}
 \end{center}
\caption{\footnotesize 
Profile of $S_\nu$ for the full range of $\nu$.
Numerical Data $N_{\rm sht}[S_\nu]$ are plotted  by  dots. 
  Dashed lines in the left panels describe $E_n[S_\nu]$ 
given in Sec.\ref{sec:TCE}.
$\widehat P_3[S_\nu]$ plotted by a solid line  and 
$\widehat P_1[S_\nu]$ plotted by a dashed-line  are defined 
in Sec.\ref{sec:PadeandLarge}.}\label{fig:Snu}
\end{figure}

 This quantity with $\nu>0$ satisfies 
\begin{eqnarray}
  0\quad <\quad S_\nu \quad < \quad\frac{ m^2}{2} \int \frac{d^2x}{2\pi} (1-e^{-\psi}) =\nu,\label{eq:SInequality}
\end{eqnarray} 
and according to Eq.(\ref{eq:vanishing}) and Eq.(\ref{eq:dpsidnu0}) we find 
\begin{eqnarray}
 \lim_{\nu\to 0}S_\nu=
\lim_{\nu\to 0}\frac{d S_\nu}{d \nu}=0,\quad 
\lim_{\nu\to 0}\frac{d^2S_\nu}{d \nu^2}=2.
\end{eqnarray}
Thanks to Eq.(\ref{eq:psiinequality2}) we find that $S_\nu$ is also
an increasing function with respect to $\nu$ and 
according to the profile of $S_\nu$ shown in Fig.\ref{fig:Snu}
 an `energy' per an unit winding number
$S_\nu/\nu$ is also  an increasing function with respect to $\nu$,
and this property gives  
\begin{eqnarray}
 S_{\nu_1+\nu_2}>S_{\nu_1}+S_{\nu_2}.\label{eq:SInequality2}
\end{eqnarray}
This inequality 
is consistent with the well known property of type I$\!$I (type I) vortices, 
that is, intervortex forces are repulsive (attractive) 
for the coupling $\lambda>1(\lambda<1)$\footnote{
It is natural to expect the following inequalities 
on values of total energies $E_k$ for axially-symmetric
vortex-solutions,
\begin{eqnarray}
E_{k_1+k_2}\gl  E_{k_1}+E_{k_2}\quad {\rm for~} \lambda \gl 1,
\label{eq:EnergyIniquality}
\end{eqnarray}
which induces the inequality (\ref{eq:SInequality}).
To the best of our knowledge, there is no known mathematical
proof for these inequalities although they are quite reasonable.}.

\subsection{Numerical Data}
We numerically calculate values of $C_\nu, D_\nu, S_\nu$ 
in most of the range of $\nu$ as  $\nu=1/20, 1/10,\cdots,500,1000$ 
using mainly the shooting method. 
These data are listed in Table.\ref{tab:NData}.
\begin{table}[t]
\begin{center}
 \begin{tabular}{|c||c|c|c|c|}
 $\nu$&$C_\nu$ & $D_\nu$ & $R_{\rm in}$& $S_\nu$
\\ \hline
1/20&0.05152300&0.007221252&1.155375 &0.002320344$^\star$
\\
1/10&0.1061386&0.01714170&1.186986&0.008668711$^\star$
\\
1/5&0.2249350&0.04429221&1.247899&0.03070642$^\star$
\\
1/2&0.6633334&0.1736933&1.415364&0.1444002$^\star$
\\
 1&1.707864&0.5053608&1.657584&0.4153533
\\
 2&5.336582&1.443305&2.057831&1.085081
\\
 3&11.86421&2.615596&2.391367&1.832041
\\
 4&22.61080&3.948209&2.683313&2.619544
\\
 5&39.31961&5.402536&2.946174&3.432922
\\
10&317.5504&13.88300&4.008030&7.704638
\\
20&5424.053&34.27687&5.550253 &16.68079
\\
50&1284274.&107.9305&8.659094&44.65765
\\
100&5.455139$\times 10^8$&250.0538&12.18905&92.38242
\\
200 & 2.607156$\times 10^{12}$& 
568.9475 & 17.19704 & 189.1678
\\
500&4.568733$\times 10^{19}$&1650.717&27.15154&482.7929
\\
1000 & 6.065189$\times 10^{27}$ & 3647.519 & 38.37932& 975.6104
\end{tabular}
\caption{\footnotesize 
Numerical Data of $C_\nu$, $D_\nu (R_{\rm in})$ and $S_\nu$.
All data are sufficiently stable values and we double-checked them 
except for data added stars. }
\label{tab:NData}
\end{center}
\end{table}
We will denote these data as $N_{\rm sht}[C_\nu], N_{\rm sht}[D_\nu]$
and $N_{\rm sht}[S_\nu]$ for $C_\nu,D_\nu,S_\nu$ respectively. 
In Sec.\ref{sec:PadeandLarge}, we use these data as  references 
to show how the winding-number expansion introduced in
Sec.\ref{sec:TCE} works well. 
The other purpose of this subsection is to settle the problem 
on the numerical value of $C_1$.   
We need, therefore, numerical calculations with  high accuracy. 
To show accuracy of our numerical data to readers, 
let us enter into details of the numerical calculations we performed.

Note that there exist two kinds of strategies in the shooting method
and we observe a big difference in usability between them.
We calculate numerical solutions of $\psi$ in a region $\{
r\,|\epsilon\le r \le  L\}$ where we set $m=1$ and 
take $\epsilon=10^{-2n+1}$ and 
$L=2\sqrt{\nu}+p \log 10 $ with $p,n= 8\sim 9$ referring to the flux
size $R_{\rm flux}$ given in Eq.(\ref{eq:Rflux}).
The first strategy is to take $r=\epsilon$  as the initial point
 of the calculation  and fine-tune the parameter $D_\nu$ so that 
a profile of $\psi$ satisfies the boundary condition at $r=L$ and 
read $C_\nu$ from a profile of $\psi$ at $r=L$.
Since the initial conditions are given by a pair $\{\nu, D_\nu\}$, 
an incorrect pair  always makes  
a profile function blow up  at large $r$.
The second one is to take $r=L$ as the initial point  
and fine-tune the parameter $C_\nu$ so that 
$\nu= -(r\psi'/2)$ at $r=\epsilon$ and read $D_\nu$ at $r=\epsilon$.
In this strategy the profile function is controlled by the only one 
initial parameter $C_\nu$ which is related to $\nu$ in one-to-one  
correspondence thanks to $dC_\nu/d\nu>1$. With the sufficiently large $L$, 
therefore, a profile function with an arbitrary $C_\nu$ always  gives a 
certain solution corresponding to a certain $\nu$, 
 without the profile blowing up,  
and thus this strategy gives a function $\nu=f(C_\nu)$.
Thanks to this property, it is easy 
to create a computer program for tuning $C_\nu$
automatically with a given $\nu$ and arbitrary precision. 
We take the second strategy in this paper although the first strategy was
taken\footnote{
He stated there as ``Hence, all numerical solutions blow up at large
$r$, and even though $a_1$ and $b_2$ were tuned to six decimal places,
the Runge-Kutta algorithm could not shoot beyond $r=10$.''
}
 in Speight's paper \cite{Speight:1996px}.

\begin{figure}[h]
 \begin{center}
  \includegraphics[width=5cm]{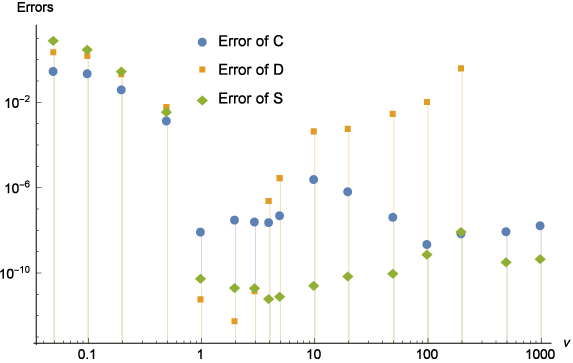}\qquad
  \includegraphics[width=6cm]{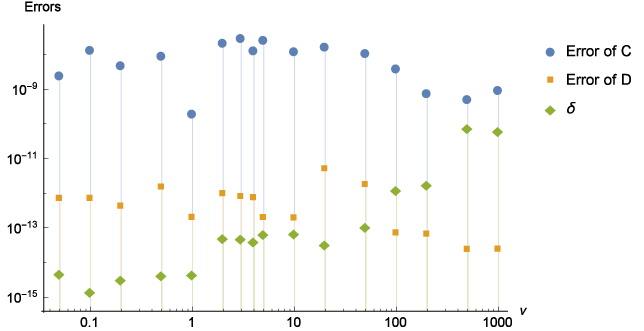}
 \end{center}
\caption{\footnotesize Estimated numerical errors:
The left panel plots  errors of numerical data calculated 
by the relaxation method from those calculated by the shooting method as,
 $|N_{\rm rlx}[X]/N_{\rm sht}[X]-1|$ with $X=C_\nu,D_\nu,S_\nu$. 
The right panel plots errors of numerical data in terms of  the shooting
 method itself as, 
$|N'_{\rm sht}[C_\nu]/N_{\rm sht}[C_\nu]-1|, 
|N'_{\rm sht}[D_\nu]/N_{\rm sht}[D_\nu]-1|$ and $\delta$.}
\label{fig:Errors}
\end{figure}

As we explained above, numerical data $N_{\rm sht}[C_\nu], N_{\rm
sht}[D_\nu]$ for $C_\nu, D_\nu$ are directly obtained. 
To double-check those data, we also use the integral formulas 
Eq.(\ref{eq:Cintegral}) and Eq.(\ref{eq:Dintegral}) for
$C_\nu, D_\nu$ respectively,
to obtain different data $N_{\rm sht}'[C_\nu], N_{\rm sht}'[D_\nu]$.
We regard 
$|N'_{\rm sht}[X]/N_{\rm sht}[X]-1|$ with $X=C_\nu,D_\nu$,
as errors of these data and plot them in the right panel of Fig.\ref{fig:Errors}.  
For instance,  we obtain  as double-checked numbers, 
\begin{eqnarray}
 N_{\rm sht}[C_1]&=&1.707864175\nn
 N_{\rm sht}[D_1]&=&0.505360825378 \label{eq:CDData}
\end{eqnarray}
for $\nu=1$ and  the numerical 
data listed in Table.\ref{tab:NData} have been  double-checked in this sense.
Therefore we conclude that 
the numerical result $C_1=1.7079$ given  by de Vega and Schaposnik is correct.
Thanks to the non-trivial identity in Eq.(\ref{eq:Derrickformula}), 
we can estimate accuracy of the profile functions itself  by calculating
the following quantity
\begin{eqnarray}
 \delta =\left|\frac1{\nu(\nu+2)}\left\{\int_\epsilon^L  dr r N[\psi] +2
				 N[C_\nu] \int_L^\infty dr r K_0(r) \right\}-1    \right|,
\end{eqnarray}
and we plotted this in the right panel of Fig.\ref{fig:Errors}.
Note that we observe that  the precision of $N_{\rm sht}[C_\nu]$ 
generally  get worse than those of $\delta, D_\nu$ as shown in
Fig.\ref{fig:Errors}. 
The precision of calculations in Speight's paper 
seems to be less than six digits
and we guess that his result $C_1\approx 1.683$ has  an error of ${\cal
O}(10^{-2})\sim {\cal O}(10^{-3})$ which is consistent with 
the other numerical results including ours. 

We obtain also a stable numerical value of $S_1$ with long digits 
\begin{eqnarray}
 N_{\rm sht}[S_1]=0.4153533072562,
\end{eqnarray}
by the shooting method.
To perform double check of the values of $S_\nu$, we also use 
 the relaxation method as the other numerical calculation. 
In the relaxation method, we introduce a relaxation time $\tau$ and 
extend $\psi(\vec x)$ to be dependent on $\tau$, 
$\psi=\psi(\vec x,\tau)$, and modify  the equations of motion  
by adding a friction term $\partial \psi/\partial \tau$ 
with an appropriate signature. 
With an appropriate initial function of $\psi$, $\psi(r,\tau=0)= 2\nu K_0(r)$ for
instance, this friction term defines the 
time evolution of $\psi$ and decreases an
`energy' of this system defined in Eq.(\ref{eq:KEnergy}).
In principle, therefore, the true solution could be obtained
with an infinite $\tau$ as 
$\psi(r)=\lim_{\tau \to \infty}\psi(r,\tau)$. 
As larger $\tau$, we will get better accuracy in many cases. 
In reality, beyond a certain finite $\tau$,
 we observe stability of values of the observables with small noises,
since those accuracy can not be better than  the calculation accuracy.
For instance we stopped the time evolutions at $\tau\approx 4\times 10^4$.  
The relaxation
 method is convenient and powerful to solve  
(simultaneous) nonlinear (partial) differential equations numerically. 
We need no fine-tuning of any parameters there.
In the simple system we are considering, however, 
the shooting method is more powerful to get precision.
Generally speaking, 
numerical data $N_{\rm rlx}[X]$ for $X=C_\nu,D_\nu,S_\nu$ calculated by the relaxation
method get worse precision as shown as Fig.\ref{fig:Errors}.
We find $|N_{\rm rlx}[S_1]/N_{\rm sht}[S_1]-1|\approx 5\times 10^{-11}$
which is guessed to be mainly an error of $N_{\rm rlx}[S_1]$.
We also get $N_{\rm rlx}[C_1]=1.707864188\dots $ and 
$N_{\rm rlx}[D_1]=0.5053608253753\dots $ again.

\section{Small Winding-Number Expansion}\label{sec:TCE}

In the paper\cite{de Vega:1976mi}, 
de Vega and Schaposnik calculated $C_1$ and $D_1$ by a semi-analytical study. Their strategy was
essentially as follows.
Let us divide the integrals in Eq.(\ref{eq:Dintegral}) and
Eq.(\ref{eq:Cintegral}) as 
\begin{eqnarray}
 \int_0^\infty  =\int_0^b  +\int_b^\infty  ,\quad {\rm with~} b\approx R_{\rm BPS}.
\end{eqnarray}
The former integral is calculated by inserting 
the expansion Eq.(\ref{eq:rexp}) which depends on $D_\nu$ and 
the latter  is calculated by the expansion
Eq.(\ref{eq:asymptoticexp}) which depends on $C_\nu$. Then  we obtain 
simultaneous equations for $C_\nu$ and $D_\nu$, and thus,
 approximate the values of $C_\nu$, $D_\nu$ as their solution.

In this section we will give a different expansion of the solution $\psi$
 using Eq.(\ref{eq:psiIntegral2}) and 
calculate them more straightforwardly and more systematically. 

\subsection{$\nu$-expansion  of the vortex function $\psi$} 
In the normal case, we can not define an expansion of $\psi$ 
with respect to the winding number as a topological quantum number.
In the previous section, we relax the winding number $\nu$ from an integer
to a real number and 
assume smoothness at $\nu=0$, and thus,
we can consider a Taylor expansion of the solution for $\psi$ 
with respect to the winding number as, with Eq.(\ref{eq:vanishing})
\begin{eqnarray}
 \psi=\sum_{n=1}^\infty \nu^n \psi_n.
\end{eqnarray}
Since the approximate solution $E_1[\psi]=\nu \psi_1$ in
 Eq.(\ref{eq:E1}) 
 satisfies the boundary conditions Eq.(\ref{eq:boundcon})  and 
has the same asymptotic form as  Eq.(\ref{eq:asymptotic})
for an arbitrary $\nu$, 
we expect that the following finite series of order $n$ 
\begin{eqnarray}
 E_n[\psi]\equiv \sum_{m=1}^n \nu^m\,\psi_m
\end{eqnarray}
gives a good approximation and  becomes better as the larger order $n$.
Here, 
a higher-order coefficient $\psi_n$ for $n\ge 2$ can be 
sequentially calculated by 
expanding the integral equation in Eq.(\ref{eq:psiIntegral}), or 
Eq.(\ref{eq:psiIntegral2}) for the axially symmetric case,
 with the first approximant $E_1[\psi]$, as
\begin{eqnarray}
 \psi_n (\vec x)
=m^2 \int \frac{d^2y}{2\pi}G(\vec x-\vec y)\sigma_n(\vec y),\quad
 \sigma[\psi]=\sum_{n=2}^\infty
  \nu^n \sigma_n 
\end{eqnarray}
where expansion coefficients 
$\sigma_n=\sigma_n(\vec x)$ 
in the interaction terms $\sigma[\psi]$ are 
\begin{eqnarray}
\sigma_2=\frac12 \psi_1^2,\quad \sigma_3=-\frac16
 \psi_1^3+\psi_1\psi_2,\quad \cdots .
\end{eqnarray}
Let us call this Taylor expansion a 
{\it small winding-number expansion }, or simply, a $\nu$-expansion. 
Note that in this expansion the winding number $\nu$ is fixed 
and higher order corrections have no logarithmic singularity as
\begin{eqnarray}
 \lim_{r\to 0}r \frac{d \psi_n}{dr}=0  \quad {\rm for~} n\ge 2.
\end{eqnarray}  
The absence of the solution for $\nu\le -1$ shown in Eq.(\ref{eq:absence})
might indicate that a radius of convergence for the $\nu$-expansion of
$\psi$ is less than 1. In Sec.\ref{sec:PadeandLarge}, 
we will discuss that this fact is  not a big problem. 

We can perform  calculations of the $\nu$-expansion of $\psi$ 
with the familiar technic using Feynman diagrams.  
The $\nu$-expansion of $\psi(\vec x)$ is given concretely as
\begin{eqnarray}
 \psi(\vec x)&=& 2\nu \times 
\hspace{-1em}\Fymn{0.25}{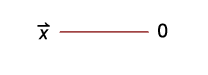}\hspace{-1em}
 +\frac12(2\nu)^2 \times \Fymn{0.2}{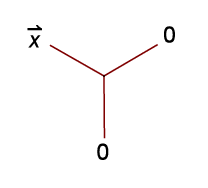}\nn
&&+ (2\nu)^3 \times \left\{
\frac12 \Fymn{0.20}{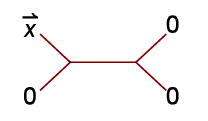}-\frac16 \Fymn{0.15}{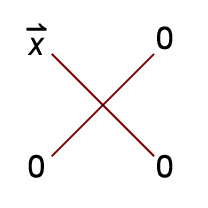}\right\}\nn
&& +(2\nu)^4\times \left\{
\frac12 \Fymn{0.20}{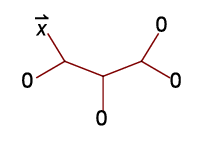}
+\frac18 \Fymn{0.20}{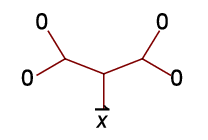}
\right.\nn
&&\qquad \left. 
-\frac16\Fymn{0.20}{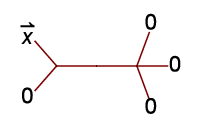}
-\frac14\Fymn{0.20}{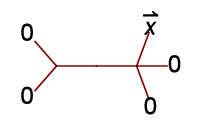}
+\frac1{24} \Fymn{0.15}{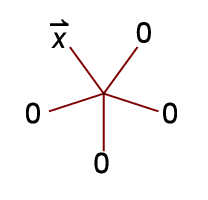}
\right\}\nn
&&+{\cal O}(\nu^5),   \label{eq:psiexpansion}
\end{eqnarray}
 using conventions for Feynman diagrams, 
\begin{eqnarray}
G(\vec x)=K_0(m|\vec x|)=\hspace{-1em}
\Fymn{0.3}{prop}
\hspace{-2em}, \quad
m^2 \int \frac{d^2y}{2\pi}G(\vec x-\vec y)G(\vec y)^2 =
\Fymn{0.2}{psv3}.
\end{eqnarray}
Here diagrams of the order $n$ have $n$ external legs coming from 
the point-like vortex at the origin $\vec x=\vec 0$.
\subsection{$E_n[C_\nu]$}
\begin{figure}[h]
 \begin{center}
  \includegraphics[width=5.5cm]{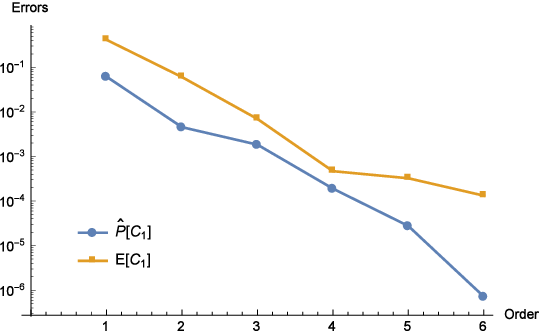}\quad
\includegraphics[width=5.5cm]{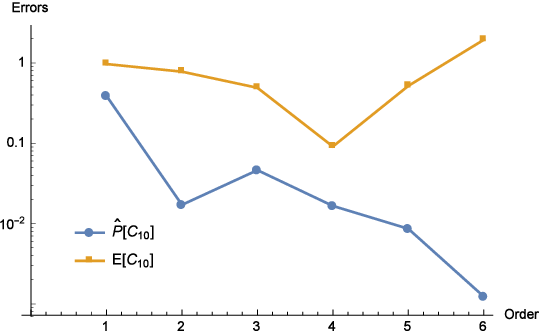}
 \end{center}
\caption{Errors of the $n$-th order approximants  $E_n[C_1]$ and $\widehat P_n[C_1]$ 
in the left panel, and $E_n[C_{10}]$ and $\widehat P_n[C_{10}]$  in the right
 panel. $\widehat P_n[C_\nu]$ will be defined in Sec.\ref{sec:PadeandLarge}}\label{fig:ErrC1C10}
\end{figure}

Let us approximate $C_\nu$  analytically by using the $\nu$-expansion,
\begin{eqnarray}
 C_\nu=\sum_{n=1}^\infty c_n \nu^n,\quad c_1=1.
\end{eqnarray}
In principle, its coefficients $c_n$ can be obtained by taking  
the $\nu$-expansions of the both sides of Eq.(\ref{eq:Cintegral}) and
inserting $\psi_n$ obtained in Eq.(\ref{eq:psiexpansion}) 
into the right hand side.
Comparing Eq.(\ref{eq:psiIntegral2}) and Eq.(\ref{eq:Cintegral}),
however, we find that  
the coefficient $c_n$ can be calculated 
by only replacing the propagator with  $I_0(m|\vec x|)$ as
\begin{eqnarray}
 \psi_n(\vec x)=\Fymn{0.20}{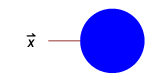}\quad \Rightarrow \quad
 c_n= \frac12 \Fymn{0.20}{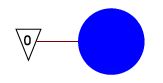} 
\end{eqnarray}
where the triangle symbol stands for
\begin{eqnarray}
 I_0(m|\vec x|)= \Fymn{0.15}{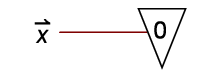}.
\end{eqnarray}
For instance, coefficients $c_2,c_3$ are calculated as  
\begin{eqnarray}
c_2&=& \Fymn{0.15}{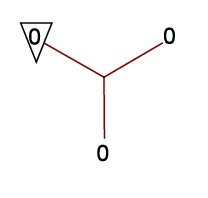}=\int_0^\infty dr r\,I_0(r)K_0(r)^2
=\frac{\pi}{3\sqrt{3}}\approx 0.604600, \nn
 c_3&=&2 \Fymn{0.18}{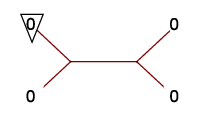}
-\frac23\Fymn{0.15}{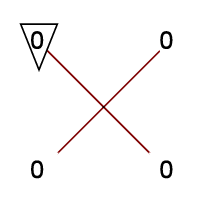} \nn
&=&2\times \frac{11}{432}\pi^2-\frac23 \times \frac{\pi^2}{16}
=\frac{\pi^2}{108}\approx 0.0913852.
\end{eqnarray}
See Appendix \ref{sec:integrals} for details.
Finally we obtain
\begin{eqnarray}
 C_\nu&=&\nu+\frac{\pi}{3\sqrt{3}}\nu^2+\frac{\pi^2}{108}\nu^3
+0.0126799 \nu^4\nn
&&\quad -0.0013557(41) \nu^5+0.000781(22) \nu^6+{\cal O}(\nu^7),
\label{eq:Cnuexpansion}
\end{eqnarray}
which gives 
a finite series $E_n[C_\nu]$ as an approximant of order $n$  
\begin{eqnarray}
 E_n[C_\nu]=\sum_{k=1}^n c_k\,\nu^k .
\end{eqnarray}
As shown in Fig.\ref{fig:ErrC1C10}, we observe that 
as the order $n$ is larger, an error of $E_n[C_1]$, that is,
$|E_n[C_1]/N_{\rm sht}[C_1]-1|$ is smaller.     
The sixth order approximant for $\nu=1$, $E_6[C_1]$, gives 
a quite nice value near to the numerical value 
$N_{\rm sht}[C_1]$ in Eq.(\ref{eq:CDData}) as
\begin{eqnarray}
 E_6[C_1]=1.70809\dots,\quad 
\left|\frac{E_6[C_1]-N_{\rm sht}[C_1]}{N_{\rm sht}[C_1]}\right|
\approx 1.0\times 10^{-4}.
\end{eqnarray}
Unfortunately the accuracy of this value is worse than that of the value 
$C_1\approx 1.7079$
given by de Vega and Schaposnik. 
According to Fig.\ref{fig:ErrC1C10} 
a radius of convergence of 
the infinite series, $\nu_{\rm c}$, is  obviously finite 
and smaller than ten, $\nu_{\rm c}<10$
 and we can not judge whether $\nu_{\rm c}$ is larger than one or not.  
In Sec.\ref{sec:PadeandLarge}, we will overcome these problems.

\subsection{$E_n[D_\nu]$}
Next, let us consider the $\nu$-expansion of $D_\nu$, 
\begin{eqnarray}
 D_\nu=\sum_{n=1}^\infty d_n \nu^n,   \quad d_1 =\log2 -\gamma. 
\end{eqnarray}
According to Eq.(\ref{eq:Dintegral}), the expansion coefficient $d_n$  
for $n\ge 2$,
is calculated by reducing diagrams in Eq.(\ref{eq:psiexpansion}) 
as,
\begin{eqnarray}
 \psi_n(\vec x)=\Fymn{0.20}{psgeneral}\quad \Rightarrow \quad
 d_n= \frac12 \Fymn{0.20}{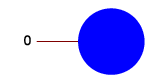}. 
\end{eqnarray}
We find therefore, by performing integrals numerically,
\begin{eqnarray}
 D_\nu&=&(\log 2-\gamma)\nu+\nu^2\times 
 \Fymn{0.15}{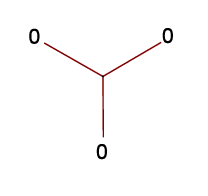}
+\nu^3\times 
\left\{2 \Fymn{0.18}{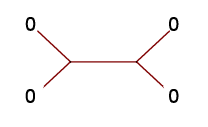} -\frac23 \Fymn{0.15}{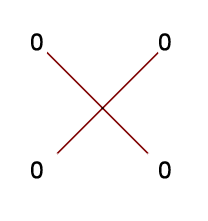} \right\}\nn
&&+\nu^4\times \left\{
5 \Fymn{0.20}{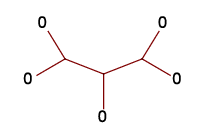} -\frac{10}3 \Fymn{0.20}{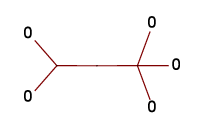}
+\frac13 \Fymn{0.20}{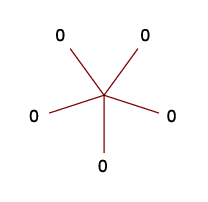} 
\right\}+{\cal O}(\nu^5)\nn
&=&0.115932 \nu+0.585977\nu^2-0.333905\nu^3+0.244999\nu^4 \nn
&& -0.196695\nu^5+0.165065(79) \nu^6+{\cal O}(\nu^7)\label{eq:Dnuexpansion}
\end{eqnarray}
and the $\nu$-expansion of $R_{\rm in}$ is also obtained as
\begin{eqnarray}
 (m R_{\rm in})^2&=&\exp\left(\frac{2D_\nu}\nu\right)\nn
&=&1.26095 +1.47777 \nu 
+0.0238675 \nu^2\nn
&&-0.030728 \nu^3 
+ 0.0300632 \nu^4 -0.02652(10)\nu^5 + {\cal O}(\nu^6). \qquad
\end{eqnarray}
Note that this quantity is known to have the lower bound $4 e^{-1}\nu$
and  the second coefficient is near to this  bound as 
$1.47777 > 4 e^{-1}=1.47152$.
Finite series 
\begin{eqnarray}
 E_n[D_\nu]=\sum_{k=1}^n d_k\, \nu^k
\end{eqnarray}
were expected to be good approximations, but we find 
their slow convergence as seen in  Fig.\ref{fig:Dnu}.  
\subsection{The $\nu$-expansion of the formula Eq.(\ref{eq:Derrickformula}) }
\label{sec:Derrickexpansion}
To check consistency of the $\nu$-expansion 
of the formula Eq.(\ref{eq:Derrickformula}), we need some unfamiliar formulas. 
There is a non-trivial identity as,   
\begin{eqnarray}
 \int d^2x \psi_n =\int \frac{d^2x\, m^2}{-\partial^2+m^2}\sigma_n=
 \int d^2x \sigma_n,
\end{eqnarray}
and using Eq.(\ref{eq:dK0dmformula}) we find
\begin{eqnarray}
 \int d^2x \psi_1 \psi_n&=& \int d^2x \sigma_n \frac{m^2}{-\partial^2+m^2
  }\psi_1=
-m^2 \int d^2x \sigma_n \frac{\partial \psi_1}{\partial m^2}.
\end{eqnarray}
Using the above formula, we also find  with $\sigma_1=4\pi \delta^2(x)/m^2$,
\begin{eqnarray}
 \int d^2x \psi_1 =\frac{4\pi}{m^2},\quad \int d^2x \psi_2
=\frac12 \int d^2x \psi_1^2
=-2\pi \frac{\partial \psi_1}{\partial m^2}\Big|_{r=0}=\frac{2\pi}{m^2}
\end{eqnarray}
and  since $\psi_3$ is a dimensionless quantity we can confirm 
\begin{eqnarray}
 \int d^2x \psi_3&=&
\int d^2x \left(-\frac16\psi_1^3 +\psi_1 \psi_2\right)
=\int d^2x \left(-\frac16\psi_1^3 -\frac12 \psi_1^2\, m^2 \frac{\partial
	     \psi_1}{\partial m^2}\right)\nn
&=&-\frac16 \frac{\partial }{\partial m^2}\left(m^2 \int d^2x \psi_1^3\right)=0.
\end{eqnarray}
 To check Eq.(\ref{eq:Derrickformula}) for more higher order, similarly 
we must need the dimensional argument again. Checking
Eq.(\ref{eq:Derrickformula}) is, therefore, tautological in this sense. 
\subsection{$E_n[S_\nu]$}
To calculate the $\nu$-expansion of $S_\nu$, 
at first we rewrite the definition of $S_\nu$ 
by inserting the identity in Eq.(\ref{eq:nusquared})
\begin{eqnarray}
S_\nu
&= & 
\frac{m^2}2\int \frac{d^2x}{2\pi}(1-e^{-\psi})^2
=\frac{m^2}2\int \frac{d^2x}{2\pi}\left(\psi^2-\psi^3+\cdots \right)
\nn
&=&\nu^2+m^2 \int \frac{d^2x}{2\pi} 
\left(\frac12(1-e^{-\psi})^2+1-e^{-\psi}-\psi \right)\nn
&=&\nu^2+m^2\int \frac{d^2x}{2\pi} 
\left(-\frac{\psi^3}3+\frac{\psi^4}4-\frac{7}{60}\psi^5
+\frac1{24}\psi^6-\frac{31}{2520}\psi^7+{\cal O}(\psi^8)\right).\qquad\quad
\end{eqnarray}
Here we canceled a $\psi^2$ term to avoid complicated and redundant
calculations such as those in Sec.\ref{sec:Derrickexpansion},
 and thus, substituting Eq.(\ref{eq:psiexpansion})
 we easily find the following expansion,\footnote{
Here a diagram of order $n$ has $n+1$ external legs.}
\begin{eqnarray}
S_\nu&=& \nu  \sum_{k=1}^\infty s_k \nu^k\nn
&=&\nu^2-\frac13(2\nu)^3\times  \Fymn{0.15}{v3}
+(2\nu)^4\times 
\left\{-\frac12 \Fymn{0.15}{v3pv3} +\frac14\Fymn{0.12}{v4}\right\}\nn
&&+(2\nu)^5\times \left\{
-\frac34 \Fymn{0.18}{v3pv3pv3}
+\frac{2}3 \Fymn{0.18}{v3pv4}
-\frac{7}{60}\Fymn{0.14}{v5}
\right\}\nn
&&+{\cal O}(\nu^6)
\end{eqnarray} 
and then,
 we obtain  by reusing 
the calculations of integrals in Eq.(\ref{eq:Dnuexpansion}) 
\begin{eqnarray}
S_\nu&=&\nu^2-1.562605\nu^3+2.73802 \nu^4\nn
&&\quad-5.05307 \nu^5+
9.59699 \nu^6-18.5461(5)\nu^7+{\cal O}(\nu^8).\label{eq:Snuexp}
\end{eqnarray}
A finite series of order $n$ for $S_\nu$ is defined as
\begin{eqnarray}
 E_n[S_\nu]=\nu \sum_{k=1}^n s_k \, \nu^k,\quad s_1=1.
\end{eqnarray}
Unfortunately we find, however, that 
these finite series  do not work as approximations even at $\nu=1$ 
as shown in Fig.\ref{fig:Snu} and it is inevitable  
to use some technique for  obtaining good approximations. 

\section{Pad\'e approximations and Large $\nu$ behaviors}\label{sec:PadeandLarge}
\subsection{ The bag model for large $\nu$ }
The result of the vortex size $R_{\rm BPS}$ in Eq.(\ref{eq:Rflux}) implies that  the total magnetic flux 
of a vortex is proportional to an area occupied by the flux 
for  $\nu>0$,
\begin{eqnarray}
 \left|\int d^2x F_{12}\right|=2\pi\nu 
= \frac{m^2}2\times \pi R_{\rm BPS}^2 
\end{eqnarray}
where $m^2/2 =e^2 v^2/2$ is the maximum of the magnetic field  allowed by 
the BPS equations Eq.(\ref{eq:bps}) for $\nu>0$. 
This fact evokes the liquid droplet model of nuclear structure,
and gives an intuitive explanation in our axially symmetric case 
for the Bradlow bound~\cite{Bradlow:1990ir}, which means just that 
the area $\pi R_{\rm BPS}^2$ must be less than the total area 
if we considered a closed two-dimensional base space.

\begin{figure}[h]
 \begin{center}
  \includegraphics[width=6cm]{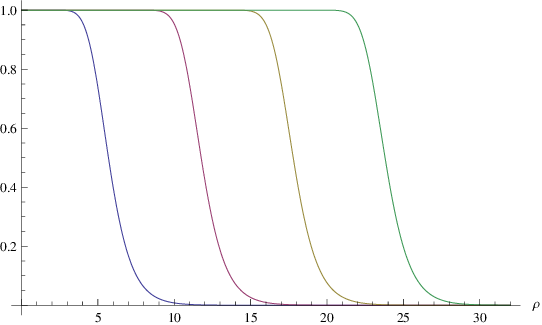}
 \end{center}
\caption{\footnotesize Configurations of the magnetic flux $-\frac{2}{m^2}F_{12}=1-e^{-\psi}$ for
 $\nu=9,36,81,144$ of which 
radiuses are estimated to be $m R=6,12,18,24$ respectively. }
\label{eq:liquidmodel}
\end{figure}

In a paper \cite{Bolognesi:2014saa}, 
the size $R_{\rm BPS}$ was obtained by a physically intuitive way using the bag
model proposed in \cite{Bolognesi:2005zr,Bolognesi:2005ty} for the large winding number $\nu$.
 In the bag model, a vortex configuration consists of 
an inside Coulomb phase, 
 the outside vacuum in the Higgs phase, 
and a thin domain-wall at $r=R$ interpolating
 their phases. In the Coulomb phase, the magnetic field takes a
 non-vanishing constant determined by  the total magnetic flux 
in Eq.(\ref{eq:topology}) with $\nu=k$, and vanishes in the vacuum. 
By omitting a thickness of the domain-wall, 
profiles of the Higgs field and the magnetic fields are approximated by
\begin{eqnarray}
 |\phi|^2=
\left\{
\begin{array}{cc}
 0& {\rm for~} r< R \\ v^2 & {\rm for~} r>R
\end{array}
\right. , \quad
|F_{12}|=\left\{ 
\begin{array}{cc}
\frac{2\nu}{R^2} &{\rm for~} r<R \\
0 & {\rm for~} r>R
\end{array}\right. ,
\end{eqnarray}
of which the total energy is calculated as
\begin{eqnarray}
T_{\rm bag}= \frac{2\pi \nu^2}{e^2R^2}+\frac{e^2v^4}8 \pi R^2 \ge \pi
 v^2 \times \nu=T_{\rm BPS}.
\end{eqnarray}
This energy is minimized just at  $R^2=4\nu/e^2v^2=R_{\rm BPS}^2$.
Actually, we numerically observe  
profiles of the magnetic field for large $\nu$ in Fig.\ref{eq:liquidmodel}.
A profile of the domain-wall is almost invariant with various values of
$\nu$.  For large $\nu$, therefore, 
 a contribution to the total energy $T_{\rm bag}$ 
form the domain-wall can be negligible. 

Since a vortex configuration for large $\nu$
drastically changes around the domain-wall at $r\approx R\gg 1/m$, 
we expect that the approximation for $r\ll R_{\rm in}$ in Eq.(\ref{eq:r0}) is 
applicable for $r=R-\epsilon <R$ with $\epsilon ={\cal O}(1/m)$ as
\begin{eqnarray}
{\cal O}(1) \approx \psi(R-\epsilon  ) 
\approx -2\nu\log \left(m(R-\epsilon)\right)+2D_\nu\approx 
2\nu \log \left(\frac{R_{\rm in}}R\right),
\end{eqnarray}
and similarly the asymptotic behavior in Eq.(\ref{eq:asymptotic}) 
is applicable for $r=R+\epsilon$
\begin{eqnarray}
{\cal O}(1)\approx  \psi(R+\epsilon ) \approx  C_\nu K_0(m(R+\epsilon) ) \approx C_\nu \sqrt{\frac{\pi}{2mR}} e^{-m
  R}.  
\end{eqnarray}
Inserting $R=R_{\rm BPS}=2\sqrt{\nu}/m$, these estimations give large-$\nu$ behaviors of $C_\nu$ and $D_\nu$ as
\begin{eqnarray}
 C_\nu\approx {\cal O}(1) \times \nu^{\frac14} e^{2\sqrt{\nu}},\quad 
D_\nu\approx \frac{\nu}2 \left(\log \nu + {\cal O}(1)
\right).\label{eq:CDlarge}
\end{eqnarray}
We also estimate $S_\nu$ as 
\begin{eqnarray}
 S_\nu=\frac{m^2}2\int \frac{d^2x}{2\pi}(1-e^{-\psi})^2&\approx& 
\frac1{4\pi} \times \pi (m R)^2 
+\frac1{4\pi}\times {\cal O}(1)\times 2\pi  m R+{\cal O}(R^0)\nn
&\approx&\nu -\beta\, \sqrt{\nu}+ {\cal O}(\nu^0).\label{eq:Slarge}
\end{eqnarray} 
Not that the term proportional to $\sqrt{\nu}$
comes from contribution of surface of the vortex and 
the coefficient $\beta$ must be positive due to Eq.(\ref{eq:SInequality}).
The above estimations for large $\nu$ will  become important clues to 
modify the  approximations using the $\nu$-expansion.

\subsection{(Global) Pad\'e approximations}
Let us assume that we know only a finite series of order $n$, 
\begin{eqnarray}
 E_n[F(\nu)]=\sum_{k=0}^n f_k \nu^k,
\end{eqnarray}
as a part of a certain
infinite  series $F(\nu)$ and it behaves as  almost an alternating
series like $F(\nu)=|f_0| -|f_1|\nu+|f_2|\nu^2-...$,
and it seems to have a  small radius of convergence $\nu\approx \nu_{\rm c}$. 
To get a good approximation for $\nu>\nu_{\rm c}$ with such a
series, it is powerful to use the Pad\'e approximation which 
replace the series by some rational functions, with $n=m+l$, 
\begin{eqnarray}
 E_{n}[F(\nu)]=P_{(m,l)}[F(\nu)]+{\cal O}(\nu^{m+l+1}).
\end{eqnarray}
where a Pad\'e approximant of
$F(\nu)$ is given by
\begin{eqnarray}
P_{(m,l)}[F(\nu)]=\frac{\displaystyle a_0+\sum_{n=1}^m a_n\nu^n}
{\displaystyle 1+\sum_{n=1}^l b_n\nu^n},
\end{eqnarray}
where coefficients of these rational functions 
are determined so that  they satisfies
\begin{eqnarray}
 \frac{d^k F(\nu)}{d\nu^k}\Big|_{\nu=0} 
=\frac{d^k}{d \nu^k}P_{(m,l)}[F(\nu)]\Big|_{\nu=0} \quad {\rm for~}
  k=0,1,\cdots, m+l.
\end{eqnarray}
Here the two sets $\{a_n\}$ and
$\{b_n\}$ are determined uniquely from the finite set $\{f_0, f_2,\cdots,f_{n+m}\}$.

There is arbitrariness in a choice of $(m,l)$ for the order $n$. 
The approximant $P_{(m,l)}[F(\nu)]$ behaves for large $\nu$ as
\begin{eqnarray}
 P_{(m,l)}[F(\nu)]\approx \frac{a_m}{b_l} \nu^{m-l}.
\end{eqnarray} 
Note that if we fix $p=m-l$ to remove that arbitrariness,
 then $n$ is restricted so that 
$n-p=2l$. In the case of $p=1$ for example, 
we arrange  
the Pad\'e approximants for all of the order $n$ as 
\begin{eqnarray}
 P_{(1,0)}[F(\nu)],\quad  \sqrt{P_{(2,0)}[F(\nu)^2]}, \quad  P_{(2,1)}[F(\nu)],
\quad \sqrt{P_{(4,2)}[F(\nu)^2]},\cdots.  
\end{eqnarray} 

\subsubsection{$\widehat P_n[S_\nu]$}

\begin{figure}[h]
 \begin{center}
  \includegraphics[width=5.5cm]{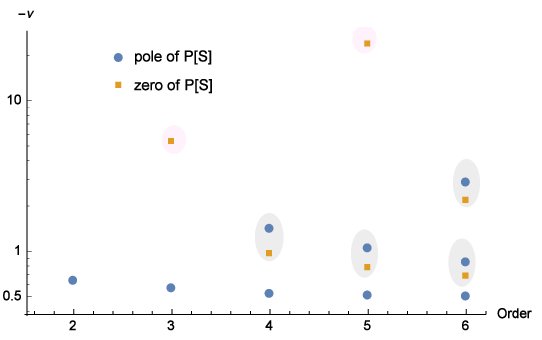}\quad
\includegraphics[width=5.5cm]{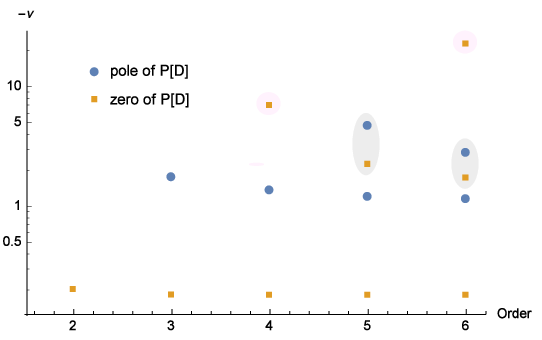}
 \end{center}
\caption{\footnotesize Poles and Zeros for $P_n[S_\nu]/\nu^2$ in the left panel, and for
 $P_n[D_\nu]/\nu$ in right panel. We observe common poles $\nu \approx
 -0.5$ for $P_n[S_\nu]$ and $\nu\approx -1$ for $P_n[D]$ and 
a common zero $\nu\approx -0.2$ for $P_n[D]$. A pair of a pole and an
 adjacent zero do not change a large-$\nu$ behavior remarkably.   }\label{fig:SDpoles}
\end{figure}

\begin{figure}[h]
 \begin{center}
  \includegraphics[width=5.5cm]{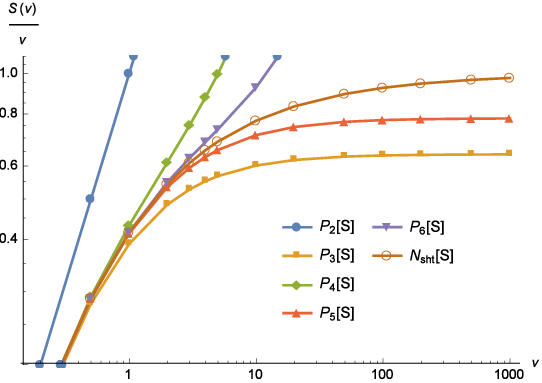}\quad
\includegraphics[width=6cm]{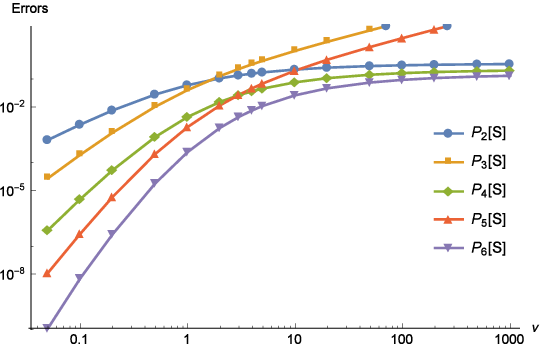}
 \end{center}
\caption{\footnotesize Profiles of $P_n[S_\nu]$ in the left panel 
and their errors from numerical data
 of $S_\nu$ calculated by the shooting method, $|P_n[S_\nu]/N_{\rm
 sht}[S_\nu]-1|$ in the right panel.}
\label{fig:SNPade}
\end{figure}

The series expansion for $S_\nu$ seems to be almost
alternative series, and according to configurations for 
the finite series $E_n[S_\nu]$ 
shown in the left panel of Fig.\ref{fig:Snu} we guess that 
the radius of convergence is around $|\nu|\approx 0.5$ which implies, for instance,  that 
the function $S_\nu$ has a singularity at $\nu\approx -0.5$.
The Pad\'e approximation can avoid such a singularity and enlarge 
the radius of convergence.      
Let us take the
following rational functions $P_n[S_\nu]$ 
with respect to $\nu$, as  Pad\'e approximants of the order $n$ for $S_\nu$,
\begin{eqnarray}
 P_2[S_\nu]&=&P_{(2,1)}[S_\nu]=\frac{\nu^2}{1 + 1.5626 \nu},\nn 
P_3[S_\nu]&=&P_{(3,1)}[S_\nu]=\frac{\nu^2 + 0.189609 \nu^3}{1 + 1.75221 \nu},\nn
P_4[S_\nu]&=&P_{(3,2)}[S_\nu]=\frac{\nu^2 + 1.05188 \nu^3}{1 + 2.61449 \nu + 1.34739
\nu^2},\nn
P_5[S_\nu]&=&P_{(4,2)}[S_\nu]=\frac{\nu^2 + 1.34536 \nu^3 + 0.0556454 \nu^4}
{1 + 2.90796 \nu + 1.86162 \nu^2},\nn
P_6[S_\nu]&=&P_{(4,3)}[S_\nu]=\frac{\nu^2 + 1.94979 \nu^3 + 0.69144 \nu^4}
{1 + 3.5124 \nu + 3.44191 \nu^2 + 
 0.814411 \nu^3}.
\end{eqnarray}
Here we have fixed arbitrariness
on choice of the Pad\'e approximants $P_{(m,n)}[S_\nu]$ so that 
all coefficients of the above are positive.\footnote{
This fact might be just by our good luck. We have no proof for
existence and uniqueness of such a choice in the all order $n$.
At least, we have to avoid zeros and poles on the positive real axis of
$\nu$ since we know $0<S_\nu<\nu$, although the arbitrariness 
remains under this restriction.} 
As a result  poles and zeros of these functions turn out to 
sit only on the negative real axis of $\nu$
as shown in Fig.\ref{fig:SDpoles} and the rational functions
$P_n[S_\nu]$ have poles around $\nu\approx -0.5$ in common.
Actually these functions give good approximations in a wider range of $\nu$ as
shown in Fig.\ref{fig:SNPade}. 
Note that  these rational functions behave as
\begin{eqnarray}
 P_{2n}[S_\nu]={\cal O}(\nu), \quad P_{2n+1}[S_\nu]={\cal O}(\nu^2),\quad {\rm
  for~large~} \nu, 
\end{eqnarray}
and $P_{2n}[S_\nu]$ give comparatively 
good approximations even for large $\nu$. This property can be understood
if we take account of the behavior of $S_\nu$ for large $\nu$ shown in 
Eq.(\ref{eq:Slarge}). Extra zeros $\nu\approx -5.27$ of
$P_3[S_\nu]$ and  $\nu \approx-23.4$ of $P_5[S_\nu]$ shown in
Fig.\ref{fig:SDpoles} can be regarded as disturbances for large-$\nu$ behaviors. 

\begin{figure}[h]
 \begin{center}
\includegraphics[width=7.5cm]{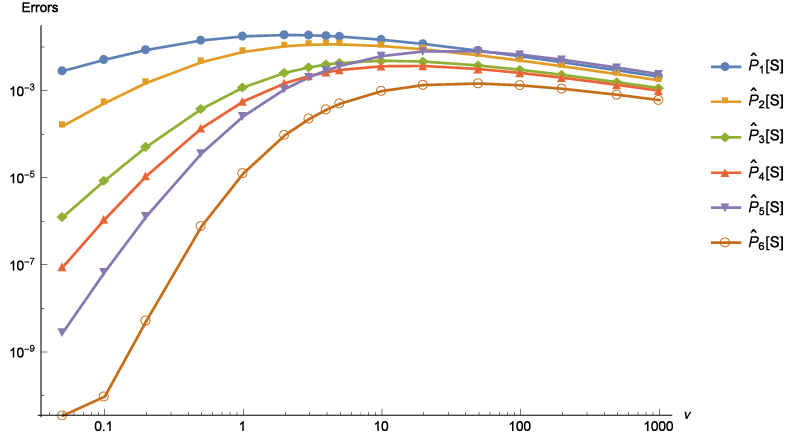}
 \end{center}
\caption{\footnotesize Errors 
$|\widehat P_n[S_\nu]/N_{\rm sht}[S_\nu]-1|$ 
of Global Pad\'e approximations $\hat
 P_n[S_\nu]$ for $S_\nu$. Distortion of a profile with $n=6$,
 at $\nu=1/20$ is consistent 
to errors of $N_{\rm sht }[S_\nu]$ itself  shown in  Fig.\ref{fig:Errors}. }
\label{fig:SGPade}
\end{figure}

\begin{figure}[h]
 \begin{center}
  \includegraphics[width=5.5cm]{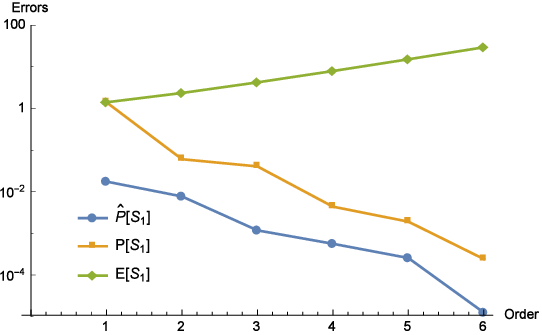}\quad
\includegraphics[width=5.5cm]{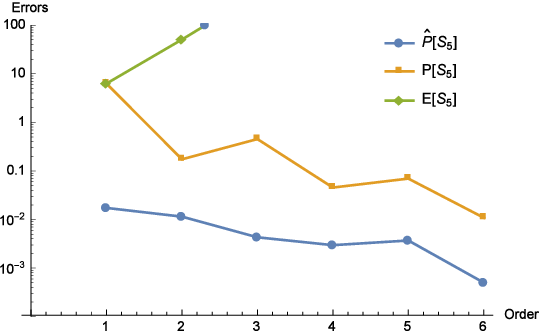}
 \end{center}
\caption{\footnotesize Errors of $S_1$ in the left panel, and $S_5$ in the right
 panel.}
\label{fig:ErrorS1S5}
\end{figure}

Let us consider the large-$\nu$ behavior  more seriously. 
The large-$\nu$ behavior in Eq.(\ref{eq:Slarge}) does not always 
mean that the function $S_\nu$ has a branch cut. 
For an example, a function $\sqrt{\nu}\tanh(\sqrt{\nu})$ has no branch cut
anywhere although it behaves $\sqrt{\nu}$ for large $\nu\in \mathbb
R_{>0}$. Here, we just assume existence of a branch cut. 
For instance, a function
\begin{eqnarray}
 \widehat P_1[S_\nu]=\nu-\nu\sqrt{\frac{1}{1+2\nu}}
\end{eqnarray}
has a branch point at $\nu=-1/2$ and  desirable behaviors as
\begin{eqnarray}
 \widehat P_1[S_\nu]=\left\{
\begin{array}{cc}
 \nu^2 +{\cal O}(\nu^3)& {\rm for~} \nu\ll 1/2 \\
 \nu-\sqrt{\frac{\nu}2}+ {\cal O}({\sqrt{\nu}}^{-1})& {\rm for~} \nu\gg 1/2
\end{array}\right.\, ,
\end{eqnarray} 
and consequently it works as a quite good approximation for the full range of
$\nu$ as shown in Fig.\ref{fig:Snu}. 
The Pad\'e approximation taking account of informations for large $\nu$ is
 called the global Pad\'e approximation~\cite{Winitzki:2003}.
Note that an expansion of the following quantity is also alternative
series due to the singularity,  
\begin{eqnarray}
 \left(1-\frac{S_\nu}\nu\right)^2&=&
1 - 2 \nu + 4.12521 \nu^2 - 8.60125 \nu^3\nn
&&\quad  + 18.0239 \nu^4 - 37.857 \nu^5 + 79.5748 \nu^6+{\cal O}(\nu^7),
\end{eqnarray}
Let us apply the Pad\'e approximation to the above series or its squared
quantity. According to Eq.(\ref{eq:Slarge}), the above quantity behaves as 
${\cal O}(\nu^{-1})$ for large $\nu$ and  this property 
fixes the arbitrariness of Pad\'e
approximants completely. Addition to $\widehat P_1[S_\nu]$ in
the above, then, we obtain the
following functions  as 
the global Pad\'e approximants of $S_\nu$,  
\begin{eqnarray}
 \widehat P_2[S_\nu]&=&\nu-\nu\sqrt[4]{\frac1{1+4\nu+3.74958\nu^2}},\nn
 \widehat P_3[S_\nu]&=&\nu-\nu
\sqrt{\frac{1 + 0.80192 \nu}{1 + 2.80192 \nu + 1.47863 \nu^2}},\nn
\widehat P_4[S_\nu]&=&\nu-\nu
\sqrt[4]{\frac{1 + 0.697034 \nu}
{1 + 4.69703 \nu + 6.53772 \nu^2 + 2.31356 \nu^3}},\nn
\widehat P_5[S_\nu]&=&\nu-\nu\sqrt{\frac{1 + 1.11774 \nu + 0.064997
\nu^2}{1 + 3.11774 \nu + 2.17527 \nu^2 + 0.0904502 \nu^3}},\nn
\widehat P_6[S_\nu]&=&\nu-\nu
\sqrt[4]{\frac{1 + 1.81492 \nu + 0.525555 \nu^2}
{1 + 5.81492 \nu + 11.5348 \nu^2 + 8.60739 \nu^3 + 1.63522 \nu^4}},\nn
\end{eqnarray}
which behave for large $\nu$ as
\begin{eqnarray}
 \widehat P_n[S_\nu]=\nu-\beta_n \sqrt{\nu}+{\cal
  O}\left(\frac1{\sqrt{\nu}}\right),
\end{eqnarray}
with coefficients for $n=1,2,\cdots$, 
\begin{eqnarray}
\{\beta_n\}=\{0.707107,0.718628,0.736437,0.740872,0.847699,0.75294,\cdots\}.
\qquad 
\end{eqnarray}
At this stage we do not know 
whether $\beta_n$ converges to a true value of $\beta$.
As we see in Fig.\ref{fig:SGPade}, the global Pad\'e approximation 
works well and  $\widehat P_6[S_\nu]$  
has a quite small errors less than $10^{-3}$ in the full range of $\nu$.
Even for small $\nu$, the global Pad\'e approximants 
$\widehat P_n[S_\nu]$ give the best result as shown in
Fig.\ref{fig:ErrorS1S5} and
the best approximant $\widehat P_6[S_\nu]$ gives 
\begin{eqnarray}
\widehat P_6[S_1]&=&0.4153585\dots, \qquad 
\left|\widehat P_6[S_1]/N_{\rm sht}[S_1]-1\right|\approx 1.3\times 10^{-5}.
\end{eqnarray}
These are the satisfactory values enough 
as results with the small winding-number expansion.
\footnote{  
We  wish, although, to modify a slow convergence
of the large-$\nu$ behavior if possible. 
Note that  a natural and probable expansion of $S_\nu$ around 
the infinity $\nu=\infty$ is
\begin{eqnarray}
 S_\nu=\nu-\beta \sqrt{\nu}+\sum_{n=0}^\infty \frac{\alpha_n}{(\sqrt{\nu})^n} 
\end{eqnarray}
although our global Pad\'e approximants $\widehat P_n[S_\nu]$ set
$\alpha_{2n}=0$. If an actual expansion has non-vanishing $\alpha_{2n}$,
 convergence of $\widehat P_n[S_\nu]$ is interfered by this feature. 
An irregular behavior of $\widehat P_5[S_\nu]$ shown in Fig.\ref{fig:SGPade} 
might be caused by this obstruction. This technical difficulty might be fatal unfortunately.  }
\subsubsection{$\widehat P_n[D_\nu]$}
\begin{figure}[h]
 \begin{center}
  \includegraphics[width=5.5cm]{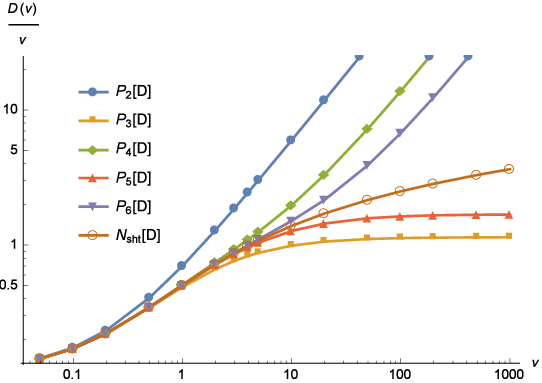}\quad
\includegraphics[width=5.5cm]{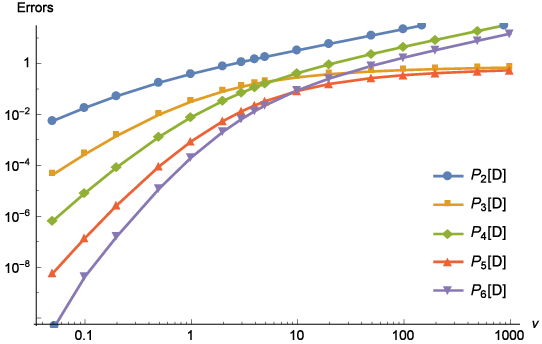}
 \end{center}
\caption{Profiles and errors of $P_n[D_\nu]$}\label{fig:DNPade}
\end{figure}

\begin{figure}[h]
 \begin{center}
  \includegraphics[width=5.5cm]{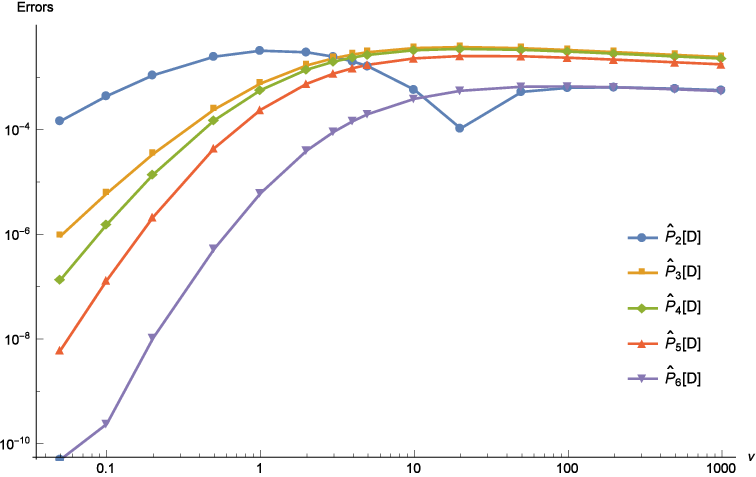}
 \end{center}
\caption{Errors in global Pad\'e approximations for $D_\nu$. }
\label{fig:ErrDGPade}
\end{figure}

\begin{figure}[h]
 \begin{center}
  \includegraphics[width=5.5cm]{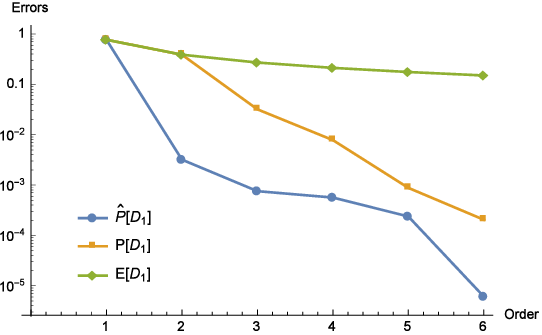}\quad
\includegraphics[width=5.5cm]{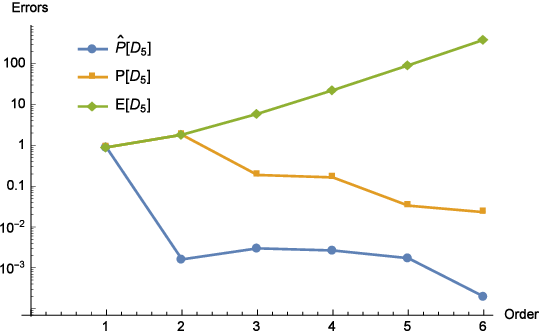}
 \end{center}
\caption{Errors of $D_1$ in the left panel, and $D_5$ in the right
 panel.}
\label{fig:ErrD1D5}
\end{figure}

The $\nu$-expansion of $D_\nu$ given in Eq.(\ref{eq:Dnuexpansion}) 
also seems to be almost an alternating series and have a
finite radius of convergence as shown in Fig.\ref{fig:Dnu}.
Hence let us consider Pad\'e approximations of $E_n[D_\nu]$.  
We can fix arbitrariness of the Pad'e approximation  
by requiring that all coefficients are positive as,   
\begin{eqnarray}
 P_3[D_\nu]&=&\frac{0.115932 \nu + 0.652038 \nu^2}{1 + 0.569826 \nu},\quad\nn
P_4[D_\nu]&=&\frac{0.115932 \nu + 0.67104 \nu^2 + 0.0960493 \nu^3}{1 +
0.733739 \nu},\quad\nn
P_5[D_\nu]&=&\frac{0.115932 \nu + 0.706900 \nu^2 + 0.297736  \nu^3}
{1 + 1.04306 \nu +  0.176257  \nu^2}\nn
P_6[D_\nu]&=&
\frac{0.115932 \nu + 0.728018  \nu^2 + 0.419974 \nu^3 + 0.0174966 \nu^4}
{1 +  1.22522  \nu + 0.309917  \nu^2},
\end{eqnarray}
which have a pole $\nu\approx -1$ in common as seen in
Fig.\ref{fig:SDpoles}.
As shown in Fig.\ref{fig:DNPade} $P_n[D_\nu]$ give  comparatively 
good approximations.
To get better approximations, 
let us apply the Pad\'e approximation not to $D_\nu$ it self, but to $
\exp(2n D_\nu/\nu)$ with $n=1,2,4$, then we obtain
\begin{eqnarray}
 \widehat P_2[D_\nu]&=&\frac{\nu}2 \log\left(\frac{0.853276 +
				      \nu}{0.676695}\right),\nn
\widehat P_3[D_\nu]&=&\frac{\nu}2 \log
\left(\frac{\sqrt{0.708551 + 1.66078 \nu + \nu^2}}{0.667557}\right),\nn
\widehat P_4[D_\nu]&=&\frac{\nu}2 \log
\left(\frac{0.654559 + 1.60982 \nu + \nu^2}{0.519101  + 0.668311 \nu}\right),\nn
\widehat P_5[D_\nu]&=&\frac{\nu}2 \log
\left(\frac{\sqrt[4]{0.511978 + 2.40006 \nu + 4.25790 \nu^2 + 3.38280 \nu^3 + \nu^4}}{0.670835}\right),\nn
\widehat P_6[D_\nu]&=&\frac{\nu}2 \log
\left(\frac{2.17939 + 5.34930 \nu + 4.17049 \nu^2 + \nu^3}
{1.72838 + 2.21671  \nu + 0.676831  \nu^2}\right).
\end{eqnarray}
These functions have the same behavior for large $\nu$ as Eq.(\ref{eq:CDlarge}),
\begin{eqnarray}
 \widehat P_n[D_\nu] \approx
  \frac{\nu}{2}\log\left(\frac{\nu}{\alpha_n}\right),\quad
  \alpha_n<e/4\approx 0.679570. 
\end{eqnarray}
Hence, as shown in Fig.\ref{fig:ErrDGPade} and Fig.\ref{fig:ErrD1D5}, 
these give quite good approximations and
errors of
 $\widehat P_6[D_\nu]$ are less than $10^{-3}$ in the full range of $\nu$.
The best approximant we obtained gives
\begin{eqnarray}
&& \widehat P_6[D_1]=0.5053639\dots,\qquad 
|\widehat P_6[D_1]/N_{\rm sht}[D_1]-1|\approx 6.1\times 10^{-6},\nn
&& 2\exp(-2 \widehat P_6[D_1])=0.727908\dots,
\end{eqnarray}
which reproduces the numerical result presented by de Vega and
Schaposnik. This value with the similar accuracy was also obtained analytically in Ref.\cite{Lozano:2000qf}.@
\subsubsection{$\widehat P_n[C_\nu]$}
The $\nu$-expansion of $C_\nu$ in Eq.(\ref{eq:Cnuexpansion}) 
gives a quite good approximation for $C_\nu$, at
least for $\nu=1$ and 
we do not know whether the radius of convergence is larger than one or not.
In this stage, therefore, it is not useful to apply the (ordinary) Pad\'e
approximation to $E_n[C_\nu]$.
Once we take account of the large-$\nu$ behavior of $C_\nu$ given in
Eq.(\ref{eq:CDlarge}), however, we notice that 
there exists a singularity at the infinity and 
we have to remove this at the first stage.
  
Let us consider the following function
\begin{eqnarray}
 {\widetilde C}_\nu\equiv \frac{\sqrt{\nu}}{2} \sinh(2\sqrt{\nu})
\end{eqnarray}
which has an infinite number of zeros on a negative real axis of $\nu$ and
 regular everywhere except for an essential singularity at the infinity. 
The  nearest next zero to the origin  is $\nu=-\pi^2/4 \approx -2.47$. 
It is, therefore,  natural to assume that a quantity 
$F_\nu\equiv (C_\nu/{\widetilde C}_\nu)^4$ has 
an infinite number of poles (and zeros) on the negative real axis of $\nu$.
Actually  we find that an expansion of $F_\nu$ gives an almost
 alternative series as,
\begin{eqnarray}
F_\nu=\left(\frac{C_\nu}{ {\widetilde C}_\nu}\right)^4 &=&
1 - 0.248268 \nu + 0.020833 \nu^2 + 0.034017 \nu^3 \nn
&&\quad - 0.0342630 \nu^4 + 
 0.0226871 \nu^5 +{\cal O}(\nu^6).\label{eq:Fseries}
\end{eqnarray}
According to Eq.(\ref{eq:CDlarge}), $F_\nu$ must behave for large $\nu$ as
\begin{eqnarray}
 F_\nu= \frac{{\rm const.}}{\nu}+{\cal O}(\nu^{-2}),\label{eq:Flarge}
\end{eqnarray}
which means that we removed the singularity at the infinity in success.
Next, let us  apply the Pad\'e approximation to the series in
Eq.(\ref{eq:Fseries}) or its squared quantity, 
satisfying the property Eq.(\ref{eq:Flarge}). 
We  obtain,\footnote{
There exists still arbitrariness on a choice of a function $\tilde
C_\nu$. We can choose, for example,  
\begin{eqnarray}
 \widetilde C_\nu= \quad \nu \cosh(2\sqrt{\nu}).
\end{eqnarray} 
However, a Pad\'e approximant of the 6-th order  with this choice 
 turn out to brake up
due to emergence of zeros or poles on the positive real axis of $\nu$.   }  
\begin{eqnarray}
 \widehat P_1[C_\nu]&=& {\widetilde C}_\nu,\quad 
\widehat P_2[C_\nu]= {\widetilde C}_\nu\sqrt[4]{\frac{1}{1+ 0.248268 \nu}},\nn
\widehat P_3[C_\nu]&=& 
{\widetilde C}_\nu\sqrt[8]{\frac{1}{1+ 0.496535 \nu + 0.143244 \nu^2}},\nn
\widehat P_4[C_\nu]&=& {\widetilde C}_\nu 
\sqrt[4]{\frac{1+0.712165 \nu}{1 + 0.960432 \nu + 0.217611
\nu^2}},\nn
\widehat P_5[C_\nu]&=& {\widetilde C}_\nu 
\sqrt[8]{\frac{1+0.600743 \nu}{1 + 1.09728 \nu + 0.441534  \nu^2 +
0.0481954 \nu^3}},\nn
\widehat P_6[C_\nu]&=&{\widetilde C}_\nu 
\sqrt[4]{\frac{1+0.709639 \nu + 0.0702914 \nu^2}
{1 + 0.957906 \nu + 0.287275 \nu^2 + 0.017348 \nu^3}}.
\end{eqnarray}
where we added $\widehat P_1[C_\nu]$ to the above although it does not
satisfy Eq.(\ref{eq:Flarge}). 
We observe the large-$\nu$ behaviors of them except for $\widehat P_1[C_\nu]$ as
\begin{eqnarray}
 \widehat P_n[C_n]\approx \omega_n \sqrt[4]{\nu}e^{2\sqrt{\nu}}
\end{eqnarray}
with coefficients
\begin{eqnarray}
 \{\omega_2,\omega_3,\cdots\}=\{{0.354169}, {0.318735}, {0.336252}, 
{0.342689}, {0.354693},\cdots\}.
\end{eqnarray}
\begin{figure}[h]
 \begin{center}
\includegraphics[width=7cm]{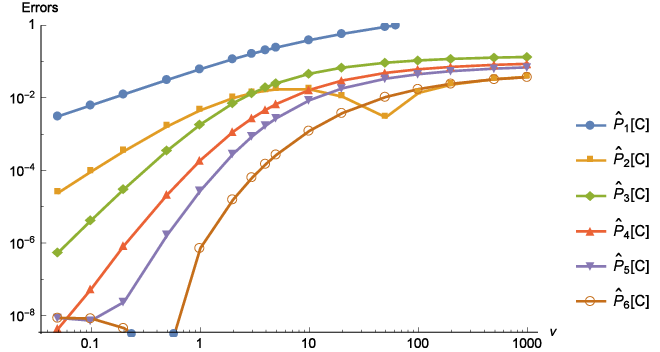}
 \end{center}
\caption{Errors of global Pad\'e approximants $\widehat P_n[C_\nu]$ for $C_\nu$.}\label{fig:CGPade}
\end{figure}
In Fig.\ref{fig:CGPade}, 
we observe that these functions give nice approximants 
in the full range of $\nu$ and 
modify  $E_n[C_\nu]$ as shown in Fig.\ref{fig:ErrC1C10}. 
Resultantly, even for $C_1$, 
we succeeded in reproducing the numerical result $C_1=1.7079$ given by de Vega and Schaposnik as
\begin{eqnarray}
 \widehat P_6[C_1]=1.7078629\dots,\quad 
\left|\frac{ \widehat P_6[C_1]}{N_{\rm sht}[C_1]}-1 \right|=7.2\times 10^{-7}.
\end{eqnarray}

%
\section{Summary and Discussion}
We considered the small winding-number expansion (the $\nu$-expansion) 
of the solution of the Taubes equation by
extending the winding number, which is a topological quantum number,
  to be a real number larger than $-1$. 
We confirmed that the $\nu$-expansion is useful to 
give good approximations of  
axially-symmetric vortex solutions in 
most of the range allowed for the winding number.
 Finally 
we found that for the scalar charge $C_1$
 the best approximate value in terms of the $\nu$-expansion  with  the
 help of the Pad\'e approximation is $\widehat P_6[C_1]=1.7078629\dots$,
which coincides with a value $N_{\rm sht}[C_1]=1.707864175$,
obtained numerically by the shooting method.
We judged that the result given by 
de Vega and Schaposnik is correct, and
Tong's conjecture giving $C_1=8^{\frac14}\approx 1.68$ from superstring theory perspective is 
incorrect as a vortex solution in the Abelian-Higgs model.
Their numerical similarity might suggest a certain universality. 

The Abelian-Higgs model of critical coupling is just the simplest toy model to test and establish 
usefulness of the $\nu$-expansion. 
The idea of the $\nu$-expansion is rather simple and
more straightforward than the strategy taken by de Vega \& Schaposnik.
As for BPS states of
vortices in further complicated systems like non-Abelian gauge-Higgs
models or of separated (parallel)
multi-vortices, therefore,
it is expected that 
the $\nu$-expansion can be
straightforwardly applied to their analytical approximations. 
  Since it is difficult to apply the shooting method 
to such complicated systems, we guess that 
the role of the $\nu$-expansion will become
more important there.   
The $\nu$-expansion is also expected to be powerful to analyze  
dependence on dimensionless parameters of solutions, 
like dependence on the number $N$ and a ratio of two gauge couplings of $U(N)=[U(1)\times
SU(N)]/\mathbb Z_N$ for an $U(N)$ vortex.
 
We expect that the $\nu$-expansion can be applied to systems
of non-critical coupling, although it might not be a straightforward
extension. Our final goal is to establish a systematic tool to study 
the dynamics of vortices  
quantitatively without taking the critical coupling limit.
Since in the $\nu$-expansion vortices are treated 
as singular particles (strings) in 
a three(four)-dimensional spacetime,
it will become possible to treat vortices of arbitrary shapes and 
discuss their dynamics analytically and quantitatively if we can consider 
such an extended $\nu$-expansion.  
\section*{Acknowledgments}
The author would like to thank Minoru Eto for a big contribution in an
early stage and Nick Manton for provision of useful information
and suggestions in a very early stage for this study. 
The author would also like to thank Hikaru Kawai, Yukinori Yasui and Shoichi Kawamoto 
for useful discussions on convergence of the expansions in various
stages.
The author is also grateful to 
 the Graduate School of Basic Sciences, University of Pisa. 
\appendix 
\section{Inequalities}
\subsection{Uniqueness of the solution}\label{sec:Uniquness}
Let us show the uniqueness of the solution  $f(\vec x)$ 
of the following $d$-dimensional partial differential equation defined
by a strictly increasing function ${\cal W}(f)$
with respect to $f$
and a source term $J$ as
\begin{eqnarray}
 -\p_i^2 f(\vec x)+{\cal W}(f(\vec x))=J(\vec x),\label{eq:DiffEq}
\end{eqnarray}  
where we require that $f(\vec x)$ vanishes at the spatial infinity.
Note  that if there exists a region $\Sigma_f$ with 
its boundary $\p \Sigma_f$  for a certain scalar function $f(\vec x)$ 
so that $f(\vec x)$ satisfies
\begin{eqnarray}
 f(\vec x) <0 ~ {\rm for~} \vec x \in \Sigma, \qquad f(\vec x)=0 
~{\rm  for ~}\vec x \in \partial \Sigma,
\end{eqnarray}
which gives $\vec n\cdot \vec \p f(\vec x)\ge 0$ with a normal vector
$\vec n$ 
of $\p \Sigma$ and then Stokes' theorem tells us  the following
inequality 
 \begin{eqnarray}
  \int_{\Sigma_f} d^dx \p^2_i f(\vec x)=\int_{\p \Sigma_f} d\vec S \cdot \vec \p f(\vec x)\ge 0.
 \end{eqnarray}
If we assume that there exist different two solutions $f_1(\vec x),
f_2(\vec x) $ for Eq.(\ref{eq:DiffEq}), then 
there exists the region $\Sigma_{\delta f}$ for a difference 
$\delta f=f_1-f_2$(or $f_2-f_1$) and 
we can derive  inconsistency as,
\begin{eqnarray}
0\le  \int_{\Sigma_{\delta f}} d^dx \p_i^2 \delta f(\vec x)=
\int_{\Sigma_{\delta f}} d^dx
\left\{{\cal W}( f_1(\vec x))-{\cal
  W}( f_2(\vec x))\right\}<0.
\end{eqnarray}
Therefore, if there exist a solution of Eq.(\ref{eq:DiffEq}), then 
it must be unique.

Furthermore,  let us consider a solution  $f(\vec x)$ 
 with  ${\cal W}(0)=0$ and $J(\vec x)\ge 0$,
\begin{eqnarray}
 -\p_i^2 f(\vec x)+{\cal W}( f(\vec x))\ge 0.
\end{eqnarray}
If there exist a region $\Sigma_f$ for this function $f$ where ${\cal W}(f)<0$, then we
find inconsistency again 
\begin{eqnarray}
 0\le \int_{\Sigma_f} d^dx \p^2_i f(\vec x)\le \int_{\Sigma_f} d^dx
  {\cal W}(f(\vec x))<0.
\end{eqnarray}
Such a solution $f(\vec x)$ must be, therefore, 
positive semidefinite everywhere.

\subsection{Sequence of sets of upper and lower bounds}\label{sec:Inequalities}
Here let us modify the inequality Eq.(\ref{eq:psiinequality}) for $\nu>0$.
\begin{eqnarray}
{\cal I}_0: \infty> \psi>0, \quad  0>P\equiv r\frac{\p \psi}{\p r}>-2\nu, 
\end{eqnarray}
to obtain a stronger set of upper and lower bounds of them.

By integrating Taubes equation and $P=r\psi'$, we find relations between
$P$ and $\psi$ using integrals as, with $Y=(r/R_{\rm in})^{2\nu}$
and setting $m=1$,  
\begin{eqnarray}
 \psi&=&\Psi[P]\equiv \lim_{\epsilon\to 0}\left\{
-2\nu \log\left(\frac{\epsilon}{R_{\rm in}}\right)+\int_\epsilon^r \frac{ds}s P(s) \right\}
=-\log Y+\int_0^r \frac{ds}s (P(s)+2\nu),\nn
 P&=&{\cal P}[\psi]\equiv -2\nu+ \int_0^r ds \,s \left(1-e^{-\psi(s)}\right).
\end{eqnarray}
Let us assume that the following set of inequalities ${\cal I}_n$ 
\begin{eqnarray}
 {\cal I}_n:  f_n^{\rm M}> \psi> f_n^{\rm m},\quad  g_n^{\rm M}> P>
  g_n^{\rm m},\qquad  {\rm for~all~} r \in \mathbb R_{>0}. 
\end{eqnarray} 
with  some given functions $f_n^{\rm M,m},g_n^{\rm M,m}$  satisfying 
\begin{eqnarray}
 \cdots  \ge f_{n-1}^{\rm M}\ge f_n^{\rm M}>f_n^{\rm m}\ge f_{n-1}^{\rm
  m}\ge \cdots \ge f_0^{\rm m}=0,\nn 
0=g_0^{\rm M}\ge \cdots  \ge g_{n-1}^{\rm M}\ge g_n^{\rm M}>g_n^{\rm m}\ge g_{n-1}^{\rm m}\ge \cdots \ge g_0^{\rm m}=-2\nu.
\end{eqnarray}
Using these inequalities,  we can construct an another set of inequalities as
\begin{eqnarray}
\Psi[g_n^{\rm M}]>\psi>\Psi[g_n^{\rm m}],\quad 
{\cal P}[f_n^{\rm M}]>P> {\cal P}[f_n^{\rm m}].
\end{eqnarray}
Therefore we obtain  a set of stronger lower and upper bounds as 
${\cal I}_{n+1}$ by
\begin{eqnarray}
g_{n+1}^{\rm M}={\rm min}\left[g_n^{\rm M}, {\cal P}[f_n^{\rm M}]
			 \right],\quad
g_{n+1}^{\rm m}={\rm max}\left[g_n^{\rm m}, {\cal P}[f_n^{\rm m}]
			 \right],\nn
f_{n+1}^{\rm M}={\rm min}\left[f_n^{\rm M}, \Psi[g_n^{\rm M}]
			 \right],\quad
f_{n+1}^{\rm m}={\rm max}\left[f_n^{\rm m}, \Psi[g_n^{\rm m}]
			 \right].
\end{eqnarray}
Consistency of these inequalities requires that 
$g_{n+1}^{\rm M}>g_{n+1}^{\rm m}$ and $f_{n+1}^{\rm M}>f_{n+1}^{\rm m}$
which reduce to, non-trivial inequalities
\begin{eqnarray}
0= g_0^{\rm M} > {\cal P}[f_n^{\rm m}],\quad \Psi[g_n^{\rm M}]> f_0^{\rm
 m}=0.
\end{eqnarray}
This couple of inequalities turns out to give lower and upper bounds for
$R_{\rm in}$ as follows.

The initial set of inequalities ${\cal I}_0$ gives
\begin{eqnarray}
 {\cal I}_1&:&\infty > \psi > {\rm max}[0,-\log Y],\quad 
\min\left[\frac{r^2}2-2\nu,0\right]>P>-2\nu,
\end{eqnarray}
and therefore we find the followings are required  
\begin{eqnarray}
0> {\max}\,{\cal P}[f_1^{\rm m}]\quad 
&\to& \quad  2\nu >\int_0^{R_{\rm in}} dr
  r\left(1-\left(\frac{r}{R_{\rm in}}\right)^{2\nu}\right)=
\frac{\nu R_{\rm in}^2}{2(1+\nu)},\nn
0<\min \Psi[g_1^{\rm M}] \quad 
&\to& \quad 
0< \frac{r^2}4 -\log Y \Big|_{r=2\sqrt{\nu}} =
\nu \log\left(\frac{R_{\rm in }^2 e}{4\nu}\right),
\end{eqnarray}
and that is, $R_{\rm in}$ must satisfy
\begin{eqnarray}
 2\sqrt{\nu+1}>R_{\rm in}>2 \sqrt{\frac{\nu}{e}},
\end{eqnarray}
otherwise a function $\psi$ can not satisfy the set of inequalities 
${\cal I}_0$ and thus blows up at large $r$. With $R_{\rm in}$ satisfying
the above set of inequalities, 
the next set of inequalities ${\cal I}_2$ can be consistently obtained as  
\begin{eqnarray}
{\cal I}_2&:& 
{\rm max}[0,-\log Y]<\psi<
\left\{ \begin{array}{cc}
\frac{r^2}4 -\log Y &  {\rm for~}r\le 2\sqrt{\nu}\\
\nu \log\left(\frac{R_{\rm in}^2e}{4\nu}\right)
 &{\rm for~}r > 2\sqrt{\nu}\end{array}\right.\,,\nn
&&
\min\left[\frac{r^2}2-2\nu,0\right]>P>
\left\{ \begin{array}{cc}
-2\nu+\frac{r^2}2\left(1-\frac{Y}{1+\nu}\right) &  {\rm for~}r\le R_{\rm
 in}\\ -2\nu+\frac{\nu R_{\rm in}^2}{2(1+\nu)} &{\rm for~}r >R_{\rm in}
	\end{array}\right.\,.\qquad\quad
\end{eqnarray}
In principle, you can calculate ${\cal I}_3,{\cal I}_4,\dots,$ sequentially
as you like. 
%
%

\section{Some Integrals}\label{sec:integrals}
Since the modified Bessel function of the second kind 
is a two dimensional Green's function, we can find the following relations
\begin{eqnarray}
&& \int d^2x K_0(m|\vec x-\vec x_1|)K_0(m|\vec x-\vec
  x_2|)\nn
&=&\frac{2\pi}{-\partial^2+m^2 }K_0(m|\vec x_1-\vec x_2|)\nn
&=&\left(\frac{2\pi}{-\partial^2+m^2 }\right)^2\delta^2(\vec
  x_1-\vec x_2)= 
-\frac{\partial}{\partial m^2}\frac{4\pi^2}{-\partial^2+m^2 }
\delta^2(\vec  x_1-\vec x_2)\nn
&=&-2\pi \frac{\partial }{\partial m^2}K_0(m|\vec x_1-\vec x_2|)
=\frac{\pi}{m}|\vec x_1-\vec x_2|K_1(m|\vec x_1-\vec x_2|) \label{eq:dK0dmformula}
\end{eqnarray}

By using the integral formulas
\begin{eqnarray}
K_0(x)=\int_0^\infty\frac{dt}{2t}e^{-\frac{x}2\left(t+\frac1t\right)},
\quad  I_0(x)=\int_0^{2\pi}\frac{d\theta}{2\pi}e^{x \cos \theta}, 
\end{eqnarray}

one can calculate the following definite integrals,
\begin{eqnarray}
 \Fymn{0.15}{Iv4}&=&\int_0^\infty dr r I_0(r)K_0(r)^3
=\int \frac{d^2x}{2\pi} e^{x_1}K_0(|\vec x|)^3\nn
&=&\int \frac{d^2x}{2\pi} 
 \frac{dt_1dt_2 dt_3}{8t_1t_2t_3}e^{x_1-(t_1+t_2+t_3)-\frac{(x_1^2+x_2^2)}{4}\left(\frac1{t_1}+\frac1{t_2}+\frac1{t_3}\right)}\nn
&=&\int_0^\infty \frac{dt_1 dt_2 dt_3}{4t_1t_2t_3
\left(\frac1{t_1}+\frac1{t_2}+\frac1{t_3}\right)}e^{-(t_1+t_2+t_3)+\left(\frac1{t_1}+\frac1{t_2}+\frac1{t_3}\right)^{-1}}\nn
&=&\frac14 \int_0^\infty\frac{du_1du_2}{(1+u_1)(1+u_2)(u_1+u_2)} = \frac{\pi^2}{16} 
\end{eqnarray}
with $t_1=s u_1,t_2=s u_2,t_3=s$,
\begin{eqnarray}
 \Fymn{0.2}{Iv3pv3}&=&\int \frac{d^2x d^2y}{4\pi^2} I_0(|\vec x|)K_0(|\vec x|)K_0(|\vec
  x-\vec y|)K_0(|\vec y|)^2 \nn
&=&\frac14\int_0^\infty
\frac{dt_1dt_2dt_3dt_4}{t_1t_2+(t_1+t_2)(t_3+t_4)}
e^{-(t_1+t_2+t_3+t_4)+\left(\frac1{t_4}+\frac1{t_3+(1/t_1+1/t_2)^{-1}}\right)^{-1}}\nn
&=&\frac{11\pi^2}{432},
\end{eqnarray}
\begin{eqnarray}
 \Fymn{0.15}{v3}&=&\int_0^\infty dr r K_0(r)^3
=\int_0^\infty \frac{dt_1 dt_2 dt_3}{4t_1t_2t_3
\left(\frac1{t_1}+\frac1{t_2}+\frac1{t_3}\right)}e^{-(t_1+t_2+t_3)}\nn
&=&\frac14 \int_0^\infty\frac{du_1du_2}{(1+u_1+u_2)(u_1+u_2+u_1u_2)}\nn 
&=&\frac{1}{36}\left\{
\psi^{(1)}\left(\frac13\right)
+\psi^{(1)}\left(\frac16\right)-\frac{8\pi^2}3
\right\}\approx 0.585977
\end{eqnarray}
where $\psi^{(1)}(x)=d^2\log\Gamma(x)/dx^2$ is the digamma function, and
\begin{eqnarray}
 \Fymn{0.12}{v4}=\int_0^\infty dr r K_0(r)^4=\frac{7}8 \zeta(3)\approx 1.051800.
\end{eqnarray}

\end{document}